\documentclass[fleqn,10pt]{SelfArx} 

\usepackage[english]{babel} 

\usepackage{lipsum} 

\definecolor{color1}{RGB}{0,0,90} 
\definecolor{color2}{RGB}{0,20,20} 

\usepackage{hyperref} 
\hypersetup{hidelinks,colorlinks,breaklinks=true,urlcolor=color2,citecolor=color1,linkcolor=color1,bookmarksopen=false,pdftitle={Title},pdfauthor={Author}}

\def\norm#1{\left |#1\right |}

\newcommand{\snot[1]}{{\scriptstyle \times 10}^{#1}}
\newcommand{\ustar}{u_{2_*}^{\phantom{3}}\!}

\JournalInfo{Accepted Manuscript - International Journal of Non-Linear Mechanics} 

\Archive{} 

\PaperTitle{A study of temporary captures and collisions in
  the Circular Restricted Three-Body Problem with normalizations of
  the Levi-Civita Hamiltonian} 

\PaperShortTitle{Captures and collisions in
  the CRTBP with normalizations of
  the Levi-Civita Hamiltonian} 

\Authors{Roc\'io I. Paez\textsuperscript{1}, Massimiliano Guzzo\textsuperscript{1}*} 
\affiliation{\textsuperscript{1}\textit{Dipartimento di Matematica ``Tullio Levi-Civita'', Universit\`a degli Studi di Padova, Italy}} 
\affiliation{*\textbf{Corresponding author}: guzzo@math.unipd.it} 

\newcommand{\keywordname}{Keywords} 

\Abstract{The dynamics near the Lagrange equilibria $L_1$ and $L_2$ of
the Circular Restricted Three-body Problem has gained attention in the
last decades due to its relevance in some topics such as the temporary
captures of comets and asteroids and the design of trajectories for
space missions. In this paper we investigate the temporary captures
using the tube manifolds of the horizontal Lyapunov orbits originating
at $L_1$ and $L_2$ of the CR3BP at energy values which have not been
considered so far. After showing that the radius of convergence of any
Hamiltonian normalization at $L_1$ or $L_2$ computed with the
Cartesian variables is limited in amplitude by $\norm{1-\mu-x_{L_1}}$
($\mu$ denoting the reduced mass of the problem), we investigate if
regularizations allow us to overcome this limit.  In particular, we
consider the Hamiltonian describing the planar three-body problem in
the Levi-Civita regularization and we compute its normalization for
the Sun-Jupiter reduced mass for an interval of energy which overcomes
the limit of Cartesian normalizations. As a result, for the largest
values of the energy that we consider, we notice a transition in the
structure of the tubes manifolds emanating from the Lyapunov orbit,
which can contain orbits that collide with the secondary body before
performing one full circulation around it. We discuss the relevance of
this transition for temporary captures.}

\begin{document}

\flushbottom 

\maketitle 


\thispagestyle{empty} 

\section{Introduction} 

\addcontentsline{toc}{section}{Introduction} 

In the last decades the close encounters of a small body with a planet
have been investigated especially in connection with the dynamics of
comets, of near-earth asteroids, and space mission
design (see \cite{conley}, \cite{Green1988}, \cite{chodas99},
\cite{jorbamas99}, \cite{Simo99}, \cite{Valsecchi99}, \cite{GJMS},
\cite{Valsecchi03}, \cite{masdemont05}, \cite{Valsecchi05},
\cite{KLMR07}, \cite{Lega11}, \cite{GomKoLoMaMasRoss},
\cite{ZCMD}, \cite{GL13}, \cite{GV13}, \cite{GL14}, \cite{celletti1},
\cite{GL15}, \cite{GL17}, \cite{LG16}, \cite{GL18}, and references
therein). A classical tool to classify a close encounter is given by
the Tisserand parameter respect to a given planet,
\begin{displaymath}
T_P= {a_P\over a}+2 \sqrt{{a\over a_P}(1-e^2)}\cos i
\end{displaymath}
($a,e$ denote the semi-major axis and the eccentricity of the small
body; $i$ denotes the inclination of the small body with respect to
the orbit of the planet; $a_P$ denotes the semi-major axis of the
planet) whose variation before and after each encounter is small. The
Tisserand parameter gives a qualitative idea of the encounter: for
increasing values of $T_P$ in a small interval around $T_P=3$, we have
the transition between the 'fast' encounters, occurring with an orbit
of the small body which is hyperbolic in the planetocentric reference
frame, and\\

{\footnotesize ©$<$2020$>$. This manuscript version is made available
  under the \\CC-BY-NC-ND 4.0 license
  http://creativecommons.org/licenses/by-nc-nd/4.0/}

\noindent
 the 'slow' close encounters which can lead to a temporary
capture of the small body.  \"Opik theory (\cite{opik}, recently
revisited in \cite{Valsecchi02}, \cite{tommei}), provided good
results in the study of the fast close encounters of comets with
Jupiter and of near-Earth asteroids with the Earth
(\cite{Valsecchi97},\cite{Valsecchi03}). The slow close encounters
are instead better studied in the framework of the dynamics generating
at the Lagrangian points $L_1,L_2$ of the Circular Restricted
Three--Body Problem, defined by the Hamiltonian
\begin{displaymath}
h= {p_x^2+p_y^2+p_z^2\over 2} + p_x y -p_y x-{\mu\over
\sqrt{(x-1+\mu)^2+y^2+z^2}}
\end{displaymath}
\vskip -0.4 cm
\begin{equation}
-{1-\mu \over \sqrt{(x+\mu)^2+y^2+z^2}}~.
\label{hamcartesianaspaziale}
\end{equation}
This Hamiltonian is written in the barycentric rotating reference
frame and with the usual units of measure for this problem: the masses
of the primaries $P_1$ and $P_2$ are $1-\mu$ and $\mu$ respectively;
their coordinates are $(x_1,0,0)=(-\mu,0,0)$, $(x_2,0,0)=(1-\mu,0,0)$
and their revolution period is $2\pi$. The Hamiltonian
  $h$ and the Tisserand parameter are related by
\begin{displaymath}
  T_P=-2h+ {\cal O}(\mu)~.
\end{displaymath}
For energy values $E > E_1$, where $E_1$ is the value of $h$
associated to the Lagrangian point $L_1$, the transits from the realm
${\cal S}$ of motions dominated by the Sun to the realm ${\cal P}$ of
motions dominated by the planet (i.e. the temporary captures) and
back, become possible (see \cite{conley}, \cite{MCK04},
\cite{AEL17}, \cite{GL18} for precise characterizations of the
different realms of motion). Respectively for $E > E_2$, where $E_2$
is the value of $h$ associated to $L_2$, the
transits from the realms ${\cal E}$ of motions which are external to
the binary system to ${\cal P}$ (i.e. the temporary captures from the
region external to the planet orbit) and back are possible.  In the
planar CR3BP (useful to study close encounters with small
inclination), \cite{conley} has shown that the orbits which perform
the transits are contained in two dimensional surfaces of the
phase-space, the so-called tube manifolds of $L_1$ and $L_2$.  These
tube manifolds are defined as follows. Let us consider for
definiteness the tube manifolds at $L_1$; for values of $E$ slightly
larger than $E_1$ there is a periodic orbit of libration around $L_1$,
the horizontal Lyapunov orbit denoted by $LL_1(E)$, whose amplitude
increases rapidly as the value of $E$ increases. All the phase-space
orbits which are asymptotic in the future (resp. in the past) to
$LL_1(E)$ form a surface which, close to $LL_1(E)$, is topologically a
2-dimensional tube extending on both right and left sides of $LL_1(E)$
called the stable tube manifold $W^s_{1}(E)$ (resp. the unstable
manifold $W^u_{1}(E)$) of $LL_1(E)$.  Analogously, for $E$ slightly
larger than $E_2$ we have the stable and unstable tube manifolds of
$LL_2(E)$. The tube manifolds of $LL_1(E)$ separate the motions which
transit between ${\cal S}$ and ${\cal P}$: for example, an orbit with
$E > E_1$ which is a temporary satellite of the Sun, when it
approaches the Lyapunov orbit $LL_1(E)$ transits to the realm ${\cal
  P}$ only if it is contained in the stable tube $W^s_1(E)$ of
$LL_1(E)$, otherwise it bounces back to the realm ${\cal
  S}$. Therefore, numerical computations of the stable and unstable
tube manifolds at various values of $E$ provide the relevant
information to understand the transit properties related to the close
encounters, e.g. to determine if a comet becomes a temporary satellite
of Jupiter. A list of comets which have been identified as potential
candidates for temporary captures can be found in \cite{CV81}.  For
this reason, the Sun-Jupiter case, with mass ratio, $\mu = \mu_{J} :=
9.537 \snot[-4]$, has received particular relevance in the literature
(e.g. see \cite{KLMR07}, \cite{LG16}, \cite{GL18}).  Also in this
paper we focus on $\mu= \mu_{J}$, while the methods
that we use can be implemented with any other value.  In particular,
we correlate a property of the tube manifolds to a property of
temporary captures which has been little considered in the literature:
the number ${\cal N}$ of revolutions performed around Jupiter during
the temporary capture and their orientation (clockwise or
counter-clockwise) measured in the rotating reference frame.  For
simplicity, we consider the right-branch of the stable tube of the
Lyapunov orbits $LL_1(E)$, and we identify three--situations:
\begin{figure*}
  \centering
  \includegraphics[width=1.6\columnwidth]{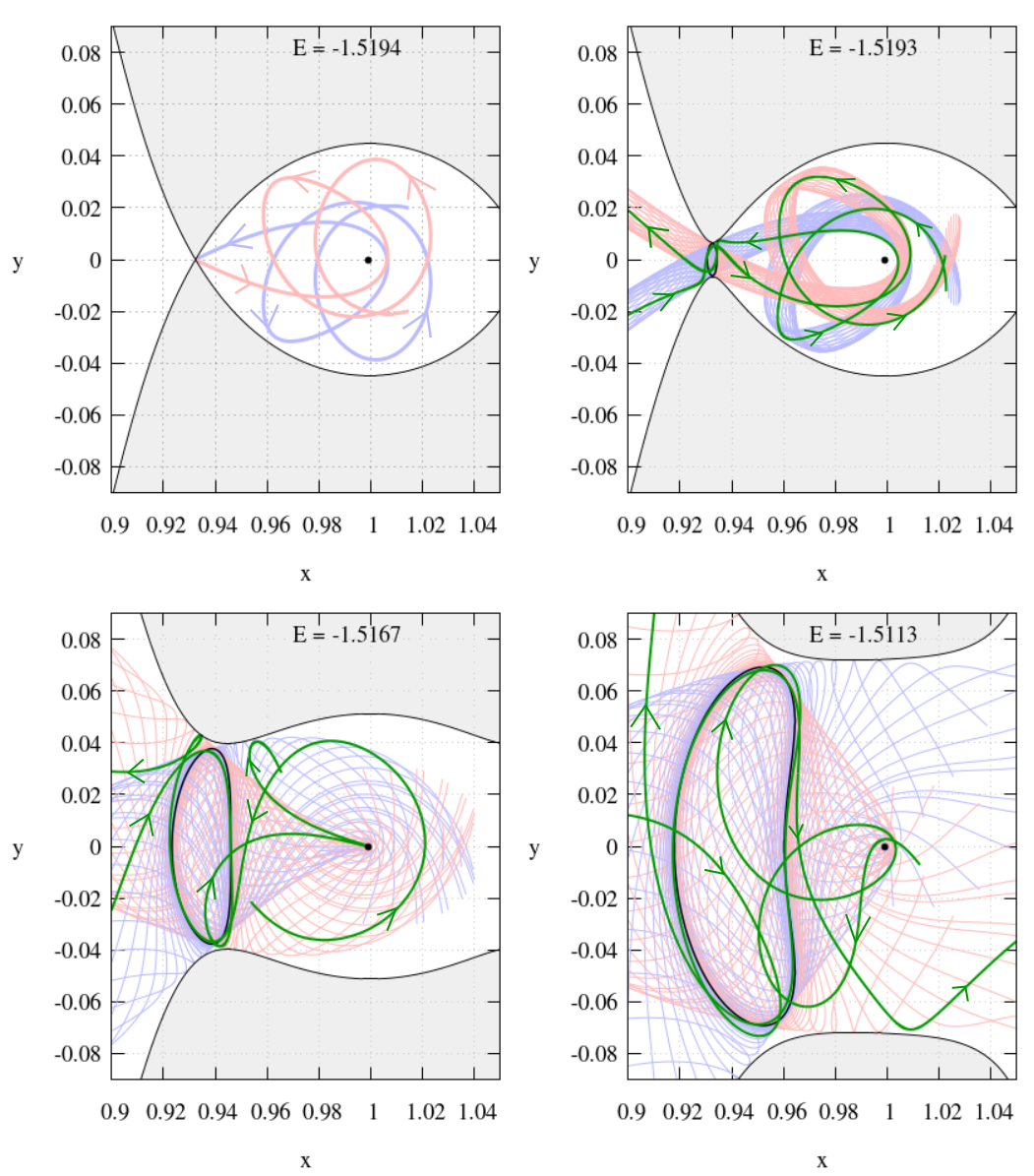}
  \caption{Lyapunov orbit $LL_1$ (black closed curves), stable tube
    manifold $W^s_1(C)$ (blue thin lines) and unstable tube manifold
    $W^u_1(C)$ (red thin lines), for four values of the energy,
    depicted in the plane $xy$ in a neighbourhood of the secondary
    body $P_2$. The shaded areas correspond to the regions of
    forbidden motion.  {\bf Top left panel:} $E = E_1 \equiv -1.5194$
    (see (i) in text); {\bf Top right panel:} $E = -1.5194$, slightly
    larger than $E_1$ (see (ii) in text); {\bf Bottom panels:} $E =
    -1.5167$ (left) and $E = -1.5113$ (right), values considerably
    larger than $E_1$ (see (iii) in text). In each case, the thick
    green lines corresponds to sample orbits from the manifolds, with the
    arrows indicating the sense of motion forward in time.}
  \label{fig:tubes}
\end{figure*}

\noindent
\begin{itemize}
\item[(i)] for $E=E_1$, the tube collapses to only one limit  
orbit (see Fig.~\ref{fig:tubes}, top-left panel);

\item[(ii)] for very small and positive values $E-E_1$, the numerical
  computation of the tube manifolds provides evidence that up to a
  fixed number ${\cal N}$ of revolutions all the orbits of the tube
  are not collision orbits, and perform \emph{the same number and type
    of revolutions}\\ \emph{around Jupiter} as the limit orbit of
  (i). As a consequence, \emph{also the orbits in the interior of the
    tube share the same properties}. In particular, since collisions
  are excluded within the ${\cal N}$ revolutions, all the orbits in
  the interior of the stable tube transit to the realm ${\cal S}$
  after the ${\cal N}$ revolutions (see Fig.~\ref{fig:tubes},
  top-right panel, for ${\cal N}=2$).

\item[(iii)] The question is what happens for increasing values of
  $E-E_1$, which corresponds to increasing amplitudes of the
  stable/unstable tubes. When these amplitude are large, we
  find that the tube manifolds become so large and stretched in
  phase-space that the criterion of using them as separatrices for the
  transit properties is no more as effective as when their amplitude
  is small.
  In this paper we propose a threshold on $E$ based on the appearance of
  peculiar orbits in the stable tube.  This orbit originates from a
  collision with Jupiter and converges to the Lyapunov orbit
  $LL_1(E)$ before performing a full revolution around the planet. In
  addition to this peculiar orbit, we find that the stable tube
  contains also orbits performing either clockwise or
  counter-clockwise revolutions (see Fig.~\ref{fig:tubes}, bottom
  panels).  \emph{The same properties are shared by the orbits in the
    interior of the tube}. In particular, due to possible collisions,
  it is no more granted that all the orbits in the interior of the
  tube will transit to the realm ${\cal S}$.

\end{itemize}

The conclusions outlined above have been obtained thanks to a method
of computation of the tube manifolds which exploits the well known
Levi-Civita regularization of the three-body problem and the method of
normalization of an Hamiltonian at a partially hyperbolic equilibrium.
The combination of the two techniques allows us to reach values of the
energy never investigated before.
  
There are several methods for
computing numerically the stable and unstable manifolds of periodic
orbits, such as the flow continuation of the local manifolds, the
parametrization method, and the recent method based on chaos
indicators.  In the first two the manifolds are developed from
analytic approximations of the local stable and
unstable manifolds (see for example \cite{KOS03}), in the latter the
manifolds are obtained as the ridges of a chaos indicator 
defined from a Hamiltonian normalization (\cite{GL14},
\cite{LG16}, \cite{GL18}).  We remark that Hamiltonian
normalizations provide not only high precision computations of the
local stable/unstable manifolds, but also of all the orbits in their
neighbourhood, suitable for astronomical applications.  For this
reason, in this paper we push the method of Hamiltonian normalization
to its limit, by implementing it in the Levi-Civita regularization of
the CR3BP.

Even if the Hamiltonian normalization method has been extensively used
to compute the tube manifolds, its limits have to be improved,
especially if one aims to extend its application to a broader interval
of energies. In the present work, we show that the use of the
Levi-Civita regularization is necessary in order to normalize the
Hamiltonian at values of the energy for which we detect the transition
from situation (ii) to (iii) described above.

In Section 2 we describe all the steps necessary to perform the
normalization of the Levi-Civita Hamiltonian. In particular, we find
that for $E> E_1$ the regularized Hamiltonian still has an equilibrium,
not corresponding to an orbit of the CR3BP. Nevertheless the Levi-Civita 
Hamiltonian can be normalized at these 'fictitious' equilibria, 
providing the Lyapunov orbit as well as their stable and unstable manifolds. 
In Section 3 we discuss the efficiency of the normal form
computations of $L_1$. In Section 4, we
show the computations of the manifolds via the normalized Levi-Civita
Hamiltonian, for a wide range of energies such that the amplitude of the 
corresponding Lyapunov orbit is larger than $\norm{1-\mu-x_{L_1}}$.  
While for small $E$ the stable tube manifold ${W}^s_1(E)$ folds around
$P_2$ exclusively in a clockwise fashion (when integrated
backwards in time for ${\cal N} =2$), for large values of $E$ 
the folding can be either clockwise or counter-clockwise. It is within
this range of energy values where we identify the transition between
the orbits discussed at ii) and iii).


\section{Hamiltonian normalizations}

Powerful methods to analyze the dynamics originating at the Lagrangian
points $L_1,L_2$ of the CR3BP rely on the Birkhoff normalizations of
the Hamiltonian with a large normalization order $N$ (\cite{GJMS},
\cite{jorbamas99}, \cite{giorgilli12}). For the planar problem, this
implies the explicit construction, for any value of an integer
parameter $N\geq 3$, of a canonical transformation
\begin{equation}
(x,y,p_x,p_y)=C_N(Q_1,Q_2,P_1,P_2)
\end{equation}
conjugating the Hamiltonian of the planar circular restricted
three--body problem
\begin{equation}\label{hamcartesiana}
  \begin{aligned}
h= & {p_x^2+p_y^2\over 2} + p_x y -p_y x-{\mu\over
  \sqrt{(x-1+\mu)^2+y^2}} \\
 & -{1-\mu \over \sqrt{(x+\mu)^2+y^2}}
  \end{aligned}
\end{equation}
to a normal form Hamiltonian\footnote{For convenience, we do not
  simplify from the Hamiltonian the constant term ${ E}_{i}$.}
{\normalsize
\begin{equation}\label{eq:hamnormalform}
H={ E}_{i}+\sum_{j= 2}^NK_j(Q_1,Q_2,P_1,P_2) +\sum_{j\geq N+1}{\cal
  R}_j(Q_1,Q_2,P_1,P_2) ~.
\end{equation}
}
The Hamiltonian in~\eqref{eq:hamnormalform} is analytic in some
neighbourhood of the Lagrange equilibrium $L_i$, represented by
$(Q,P)=(0,0,0,0)$ (the radius of the neighborhood depending on the
mass ratio $\mu$ and $N$), $K_j,{\cal R}_j$ are polynomials in 
$P,Q$ of order $j$, the polynomials $K_j$ depend on the $Q_1,P_1$ only
through the product $Q_1P_1$ and
\begin{equation}\label{eq:hamcartesian2}
K_2=\lambda Q_1 P_1+ \omega \, {P_2^2+Q_2^2\over 2}~,
\end{equation}
(alternative reduction methods can be considered, see Section 3 for
details). By neglecting the remainder terms ${\cal R}_j$ we obtain
\begin{itemize}
\item The approximated equations of the Lyapunov orbits labeled by
  $E$
\begin{equation}
{LL}_i(E)= \Big \{ Q,P:\ \ Q_1,P_1=0 ,\ \ E_i +\sum_{j=2}^N{ K}_j
(0,Q_2,0,P_2)=E \Big \}~,
\label{lli}
\end{equation}
\item the approximated equations of the (local) stable and unstable
  manifolds of ${LL}_i(E)$
\begin{equation*}
  \begin{aligned}
{ W}^s_i(E)=\Big \{ Q,P:\ \ Q_1=0 ,\ \ E_i +\sum_{j=2}^N{ K}_j
(0,Q_2,P_1,P_2)=E \Big \} , \cr { W}^u_i(E)= \Big\{ Q,P:\ \ P_1=0
,\ \ E_i +\sum_{j=2}^N{\cal K}_j (Q_1,Q_2,0,P_2)=E \Big \} ,
  \end{aligned}
\end{equation*}
\item the arcs of approximated orbits in a neighbourhood of ${
  W}^s_i(E)$, ${W}^u_i(E)$ which are scattered by $LL_i(E)$.
\end{itemize}

For increasing values of $E$, the periodic orbit ${LL}_i(E)$ has
increasing libration amplitude.  As a consequence, for suitably large
values of $E$, we expect a breakdown of the method. A natural limit
for this breakdown is given by the singularity in the gravitational
potential energy at the position of $P_2$ which, for small values of
$\mu$, is close to $L_1,L_2$.  Since the Lyapunov orbits ${LL}_i(E)$
are typically larger in the $y$ coordinate with respect to the $x$
coordinate, a dangerous complex singularity is the one located at
$x=x_{L_j}$ and $y= \pm i\norm{1-\mu-x_{L_j}}$ (for
  $\mu = \mu_J$, $y =\pm i 0.066...$). This singularity severely
  limits the validity of these methods for libration amplitudes in the
  $y$ variable of order of $\norm{1-\mu-x_{L_j}}$. We here
investigate the possibility to overcome this limit by implementing the
normalization methods using the Levi-Civita regularization
(\cite{LC1906}) on the secondary body $P_2$. Similar
approaches has been used in the past for the simpler Hill's problem,
e.g.~\cite{simo}.

\vskip 0,2 cm Following \cite{LC1906}, we first perform the
phase-space translation
\begin{equation}\label{eq:planetoXYZ}
X=x-x_{2}~, \quad Y=y~, \quad P_X=p_x~, \quad P_Y=p_y-x_{2}~~,
\end{equation}
on Hamiltonian (\ref{hamcartesiana}),
and then we introduce the Levi-Civita variables
$(u_1,u_2)$ canonically extended to the
momenta $(U_1,U_2)$
\begin{equation}\label{eq:lc_var}
  \begin{aligned}
    & X = u_1^2-u_2^2~, Y = 2u_1 u_2~, \\ &\text{\normalsize $P_X =
      \frac{U_1 u_1 - U_2 u_2}{2 |u|^2}$}~,\text{\normalsize $P_Y =
      \frac{U_1 u_2 + U_2 u_1}{2 |u|^2}$}~,
  \end{aligned}
\end{equation}
and the fictitious time $\tau$
\begin{equation}\label{eq:lc_vartau}
dt  = |u|^2 d\tau
\end{equation}
where $| {u}|^2 = u_1^2+u_2^2= \sqrt{(x-1+\mu)^2+y^2}$.  
For any fixed value of $E$ of the
Hamiltonian (\ref{hamcartesiana}), we define 
the Levi-Civita Hamiltonian
{\small
\begin{equation}\label{eq:KofE}
  \begin{aligned}
    & {\cal K}_{E}(u,U) = \\
   & =\norm{{u}}^2 \left [ h
        \Big ( X(u)+x_2,Y(u),P_X(u,U),P_Y(u,U)+x_2 \Big ) -E \right]  & \\
   & =\norm{{u}}^2 \left [ h
        \left(\text{\small $u_1^2-u_2^2+x_2$},
        \text{\small $2 u_1u_2$},
        \text{\footnotesize $\frac{U_1 u_1 - U_2 u_2}{2 |u|^2}$},
        \text{\footnotesize $\frac{U_1 u_2 + U_2 u_1}{2 |u|^2+x_2}$}
        \right) -E \right] &
    \\ &=\frac{1}{8} \left( U_1+2 \norm{{u}}^2 u_2 \right)^2 +
    \frac{1}{8} \left(U_2- 2\norm{{u}}^2 u_1 \right)^2 \\&
    -\frac{1}{2} \norm{{u}}^6 -\mu -|{u}|^2 \left(E +
    \frac{(1-\mu)^2}{2} \right) \\ &- (1-\mu) | {u}|^2 \left[
      \frac{1}{\sqrt{1 + 2 (u_1^2-u_2^2) + |{u}|^4}} + u_1^2 - u_2^2
      \right]~, & 
\end{aligned}
\end{equation}}
which is a regularization of the planar three-body problem at $P_2$. 
In fact, ${\cal K}_{E}$ is regular at $u=(0,0)$
which correspons to a collision with $P_2$. The solutions
$({u}(\tau), {U}(\tau))$ of the Hamilton equations of ${\cal K}_{E}$, 
\begin{equation}
{d u_j \over d\tau} = {\partial\over \partial U_j}{\cal K}_E(u,U)\ \ ,\ \ 
{d U_j \over d\tau} = -{\partial\over \partial u_j}{\cal K}_E(u,U)
\end{equation}
with initial conditions\footnote{The initial conditions of the
  regularized variables correspond to initial conditions of the
  original barycentric reference frame such that
  $h(x(0),y(0),p_x(0),p_y(0))=E$.} satisfying $ {u}(0)\ne 0$ and\\
${\cal K}_{E}( {u}(0), {U}(0))=0$, are conjugate, in a neighbourhood
of $\tau=0$, via Eq.~\eqref{eq:lc_var} and
\begin{displaymath}
t = \int_0^\tau \norm{u(s)}^2 ds  ,
\end{displaymath}
to solutions
$(X(t),Y(t),P_X(t),P_Y(t))$ of the three-body problem
Hamiltonian (see \cite{CG19a}). For definiteness, we describe the
normalization of the Levi-Civita Hamiltonian at the Lagrangian point
$L_1$; a similar method applies to $L_2$.
\vskip 0.2 cm
\noindent
{\bf Equilibrium points of ${\cal K}_E$.} The regularized Hamiltonian
${\cal K}_E$, for increasing values of $E\geq E_{1}$, has a family of
equilibria continued from $L_1$. Since the equilibria of the family
are characterized by $u_1,U_2=0$ and $U_1=-2 u_2^3$, we define the
fictitious equilibrium position as
\begin{equation}\label{fictitious}
(u_*,U_*):=(0,\ustar \,, -2\left(\ustar \right)^3,0)~,
\end{equation}
where $\ustar := u_{2_*}^{\phantom{3}}\!(E)$ solves the algebraic
equation \\$F(\ustar(E),E)=0$ with
\begin{equation}\label{eq:fu2e}
  \begin{aligned}
 F(&\ustar,E)  = -2\left(E + \frac{1}{2} (1-\mu)^2\right) - 3
 (\ustar)^4 \\ & +(1-\mu)\left (4 (\ustar)^2 -{2 (\ustar)^2\over
   (1-(\ustar)^2)^2}-{2\over 1-(\ustar)^2}\right ) ~.
  \end{aligned}
\end{equation}
In Appendix 1 (see paragraphs A1 and A2) we prove that equation
$F(u_{*,2},E)=0$ has a unique solution $\ustar$ for all the values of
$E$ in an interval $E\in [E_{1},E_*]$ where
\begin{equation}\label{Ecrit}
E_{*}=-{3\over 2}\left ( 1 - {4\over 3}\mu + {1\over 3}\mu^2\right ) .
\end{equation}
The critical value (\ref{Ecrit}) appears in the study of close
encounters with $P_2$ as a threshold value for considering a close
encounters fast or slow (see, for example, Section 3.1 of
\cite{GL13}). With this classification we are considering the regime
of slow close encounters. The function $\ustar(E)$ is strictly
monotone increasing, and the extremal values range from
$\ustar(E_{1})=-\sqrt{1-\mu-x_{L_1}}$ (so that
$(u_1,u_2)=(0,\ustar(E_{1}))$ corresponds to the equilibrium $L_1$) to
$\ustar(E_*)=0$ (so that $(u_1,u_2)=(0,0)$ is the collision
point). Except for $E=E_{1}$, we have ${\cal K}_E(u_*,U_*)\ne 0$,
therefore $(u_*,U_*)$ does not correspond to a
solution of the three--body problem (no equilibria of the CR3BP
different from $L_1$ exist for $y=0$ and $x\in
(-\mu,1-\mu)$). Nevertheless the equilibrium $(u_*,U_*)$ can be used
to perform a Birkhoff normalization of the Levi-Civita Hamiltonian
(\ref{eq:KofE}).  We remark that for small values of $\mu$ the energy
interval $(E_{1},E_{*})$ seems small (for instance, for $\mu=\mu_J$, 
$(E_{1},E_{*})\sim (-1.5193,-1.49809)$) but tiny
variations of the energy in this interval produce large variations in
the amplitude of the Lyapunov orbits.

In Appendix 1, paragraph A3, we show that for any $\mu\in (0,1/2)$,
and any $E\in (E_{1},E_*)$ the fictitious equilibria $(u_*,U_*)$ are
of saddle-center type, with two imaginary eigenvalues $\pm i \omega$
and two real eigenvalues $\pm \lambda$, with
$\lambda,\omega>0$. Moreover, for $a$ going to zero, both
$\lambda,\omega$ tend to zero as 
$(\ustar)^2$.

\vskip 0.4 cm
\noindent
{\bf The quadratic part of ${\cal K}_E$ expanded at $(u_*,U_*)$.} For
all $E\in (E_{1},E_*)$, the Levi-Civita Hamiltonian ${\cal K}_E$ has a
saddle-center equilibrium in $(u_*,U_*)$ and, as indicated in Appendix
1 paragraph A4, one can construct a canonical change of variables
\begin{equation}\label{eq:diagtrans}
   (u_1,u_2,U_1,U_2) = A \ (\xi, q ,\eta, p) + (u_{*,1},u_{*,2},U_{*,1},U_{*,2})
\end{equation}
conjugating ${\cal K}_E(u,U)$ to the Hamiltonian
\begin{equation}\label{eq:expham}
    {\cal K}^{(2)}_E(\xi, q ,\eta, p) =K_0+ K_{2}(\xi,q,\eta,p) + \sum_{j\geq 3}
    k_{j}(\xi,q,\eta,p)~,
\end{equation}
where $K_0={\cal K}_E(u_*,U_*)$,
\begin{equation}\label{k2}
  K_{2} =\lambda\, \xi\, \eta+ i \omega\, q\, p~,
\end{equation}
and $k_{j}(\xi,\eta,q,p)$ are homogeneous polynomials  of degree $j$
expressed as sum of monomials of the type
\begin{equation}\label{eq:monomial}
 \alpha^{(2)}_{k_1,\,l_1,\,k_2,\,l_2} \,q^{k_1}\, p^{l_1}\, \xi^{k_2}\,  \eta^{l_2}\ .
\end{equation}
Note that the couple of conjugate
variables $\xi,\eta$ is related to the hyperbolic behaviour, while the
couple of conjugate variables $q,p$ is related to the elliptic
behaviour (moreover, $q,p\in {\Bbb C}$ but, for real values of $u,U$, they satisfy
$q=-i p$).  The Hamiltonian in Eq.~\eqref{eq:expham} is the
starting point for the algorithm performing the Birkhoff
normalization. 
\vskip 0.4 cm
\noindent
{\bf Normal forms of the Levi-Civita Hamiltonian at $L_1$.} We
reproduce with the Levi-Civita Hamiltonian the normalization methods
which have been introduced in the Cartesian variables
(\cite{jorbamas99},~\cite{giorgilli12}). We perform a
normal form scheme that uncouples (up to any arbitrary order) the
hyperbolic variables $\xi,\eta$ from the elliptic variables $q,p$.
The normalization is a near to the identity canonical
  transformation ${\cal C}_N$ which conjugates the Hamiltonian
  \eqref{eq:expham} to the normal form Hamiltonian
\begin{equation}\label{normLC}
\begin{aligned}
{\cal K}^{(N)}_E & =   {\cal K}^{(2)}_E({\cal C}_N(\xi,q,\eta,p)) \\
& = 
K_0+ \lambda\, \xi\, \eta+ i \omega\, q\, p + \sum_{j=3}^{N} K_{j}(\xi,q,\eta,p) \\
 & + \sum_{j   \geq N+1} {\cal R}^{(N)}_j (\xi,q,\eta,p),
\end{aligned}
\end{equation}
where $N\geq 3$ is an integer called normalization order, 
${\cal R}^{(N)}_j$ are polynomials of order $j$ expressed as sum of monomials 
of the type:
\begin{equation}\label{eq:monomialN}
 \alpha^{(N)}_{k_1,\,l_1,\,k_2,\,l_2} \,q^{k_1}\, p^{l_1}\, \xi^{k_2}\,  \eta^{l_2}\ ,
\end{equation}
while the $K_{j}$ are polynomials of order $j$ having a specific form
defined by a certain strategy (see \cite{Meyeretal},
\cite{Sandersetal} for an introduction to polynomial normal forms). A
traditional strategy is to require that all the $K_{j}$ commute with
$K_2$. For motivations specific of the three-body
problem, such as the optimization of the computational time (which is
crucial to perform a large number $N\geq 20$ of normalizations) and due to 
the bifurcations occurring in the spatial case preventing
the integrability on the center manifold, the
reduction is usually performed with a weaker normal form. In this
paper, for the sake of comparison, we consider the three strategies:
\begin{itemize}
\item[(a)] the $K_{j}$ depend on the variables $\xi$,
  $\eta$ only through the product $\xi \eta$ (i.e.  the canonical
  transformation eliminates from the normal form every monomial of
  order smaller or equal than $N$ for which $k_2 \neq l_2$, see
  \cite{jorbamas99}). As a consequence, by neglecting
    the terms ${\cal R}_j$ (which are of order $j\geq N+1$),
    from the normal form Hamiltonian ${\cal K}^{(N)}_E$, we obtain an
    approximate normal form of type
    \begin{equation}\label{normLCa}
      K^{(N)}_E := K_0+ \lambda\, \xi\, \eta+ i \omega\, q\, p + {\cal
        J}(\xi \eta,q,p)
    \end{equation}
    where ${\cal J}$ is polynomial. From (\ref{normLCa}) one computes
    the center manifold (corresponding to $\xi,\eta=0$) as well as its
    stable (corresponding to $\xi=0,\eta\ne 0$) and unstable
    (corresponding to $\eta=0,\xi\ne 0$) manifolds.

\item[(b)] the $K_{j}$ contain only two type of
  monomials (\ref{eq:monomial}): monomials independent of $\xi,\eta$
  and dependent on the $q,p$ only through the product $q p$, as well
  as monomials at least quadratic in $\xi,\eta$ (i.e. the canonical
  transformation eliminates every monomial for which $k_2+l_2 = 1$, or
  those which simultaneously satisfy that $k_2 + l_2 = 0$ and $k_1
  \neq l_1$, see \cite{giorgilli12}). Neglecting the
    polynomials ${\cal R}_j$, we obtain an approximate normal form of
    type
    \begin{equation}\label{normLCb}
      \begin{aligned}
        K^{(N)}_E :&= K_0+ \lambda\, \xi\, \eta+ i \omega\, q\, p + {\cal
          J}_0(qp)\\ &+\xi^2 \,{\cal J}_1(\xi,q,\eta,p)+ \eta^2 \,{\cal
          J}_2(\xi,q,\eta,p) \\ &+\xi \,\eta\, {\cal J}_3(\xi,q,\eta,p)
      \end{aligned}
    \end{equation}
    where ${\cal J}_0,\ldots,{\cal J}_3$ are polynomial functions.
    From (\ref{normLCb}) 
    one computes the center manifold (corresponding to $\xi,\eta=0$), 
    but not its stable and unstable manifolds.

\item[(c)] the $K_{j}$ contain only two type of
  monomials (\ref{eq:monomial}): monomials independent of $\xi,\eta$
  and monomials at least quadratic in $\xi,\eta$ (i.e. the canonical
  transformation eliminates every monomial for which $k_2+l_2 = 1$, to
  our knowledge this strategy has been never introduced
  before). Neglecting the polynomials ${\cal R}_j$, we
    obtain an approximate normal form of type
  \begin{equation}\label{normLCc}
    \begin{aligned}
      K^{(N)}_E : &= K_0+ \lambda\, \xi\, \eta+ i \omega\, q\, p + {\cal
        J}_0(q,p) \\
      & +\xi^2 {\cal J}_1(\xi,q,\eta,p)+ \eta^2 {\cal
        J}_2(\xi,q,\eta,p) \\ &+\xi \eta {\cal J}_3(\xi,q,\eta,p)
    \end{aligned}
\end{equation}
where ${\cal J}_0,\ldots,{\cal J}_3$ are polynomial functions.  From
(\ref{normLCc}) one computes the center manifold (corresponding to
$\xi,\eta=0$), but not its stable and unstable manifolds.

\end{itemize}
While all these strategies allow one to compute the center manifolds,
only strategy (a) allows one to compute also its stable and unstable
manifolds. On the other hand, for any given normalization order $N$,
the normal form (a) has less monomial terms than the normal form (b),
which in turn has less monomial terms than the normal form (c). Thus,
the computation of the normal form (a) requires the elimination of a
larger number of monomials, and is heavier than the computation of the
normal form (b) or (c).  Therefore, if one is interested only in the
computation of the Lyapunov orbits, then the best strategy is to
adhere to the procedure that requires the minimum amount of
eliminations, i.e. (b) or (c). On the other hand, if one is interested
also in the computation of the stable and unstable manifolds
$W^s_1(E),W^u_1(E)$, and the dynamics in their neighbourhood, then one
adheres to the procedure (a).

\vskip 0.4 cm
\noindent
{\bf Algorithm for the construction of the normal forms.}
For each $N\geq 3$, the canonical transformation ${\cal C}_N$ conjugating  the 
Hamiltonian \eqref{eq:expham} to the normal form Hamiltonian (\ref{normLC})
is obtained from the composition of canonical transformations:
\begin{equation}\label{eq:totalC}
  {\cal C}_N= {\cal C}_{\chi_{N}}\, \circ \, {\cal C}_{N-1}~,
\end{equation}
where ${\cal C}_{\chi_{N}}$ is the Hamiltonian flow at time $t=1$ of a
suitable generating function defined from the coefficients of ${\cal
  K}^{(N-1)}_E$, and ${\cal C}_{2}$ is the identity.  Below we
describe the steps required for the computation of each ${\cal C}_N$
and ${\cal K}^{(N)}_E$ using the Lie series method (for an
introduction to the method, see \cite{laplatanotes},
\cite{pisanotes}). We assume that the normal form ${\cal K}^{(N-1)}_E$
and ${\cal C}_{N-1}$ are known.

\vskip 0.1 cm
\noindent
{\underline {\it Step 1:}} From ${\cal K}^{(N-1)}_E$, we compute the generating
function ${\chi_{N}}$, according to the chosen strategy:
\vskip 0.2 cm

\noindent
--\ For strategy (a)
\begin{equation}\label{eq:genfunc-a}
  \chi_N = \hspace{-0.6cm}\sum_{\substack{ {k},{l}\in {\Bbb N}^2:\\ \sum_n
      (k_n+l_n)=N,\\ k_2 \ne l_2}} \hspace{-0.3cm}
      {-\alpha^{(N-1)}_{k_1,l_1,k_2,l_2}\over i\omega
        (l_1-k_1)+\lambda (l_2-k_2)} \,q^{k_1}\, p^{l_1}\, \xi^{k_2}\,
      \eta^{l_2}~.
\end{equation}

\noindent
--\ For strategy (b)
\begin{equation}\label{eq:genfunc-b}
  \chi_N = \hspace{-0.6cm} \sum_{\substack{ {k},{l}\in {\Bbb N}^2:\\ \sum_n
      (k_n+l_n)=N,\\ k_2 +l_2 =1 \,\lor\\ k_2 +l_2 = 0 \, \land \, k_1
      \neq l_1 }} \hspace{-0.4cm} {-\alpha^{(N-1)}_{k_1,l_1,k_2,l_2}\over i\omega
    (l_1-k_1)+\lambda (l_2-k_2)} \,q^{k_1}\, p^{l_1}\, \xi^{k_2}\,
  \eta^{l_2}~.
\end{equation}

\noindent
--\ For strategy (c)
\begin{equation}\label{eq:genfunc-c}
  \chi_N =  \hspace{-0.6cm} \sum_{\substack{ {k},{l}\in {\Bbb N}^2:\\ \sum_n
      (k_n+l_n)=N,\\ k_2 + l_2 = 1}}  \hspace{-0.3cm}
      {-\alpha^{(N-1)}_{k_1,l_1,k_2,l_2}\over i\omega
        (l_1-k_1)+\lambda (l_2-k_2)} \,q^{k_1}\, p^{l_1}\, \xi^{k_2}\,
      \eta^{l_2}~.
\end{equation}
\vskip 0.1 cm

\noindent
{\underline {\it Step 2:}} We compute the canonical transformation
\begin{displaymath}
 {\cal C}_{\chi_{N}}(   \tilde \xi,\tilde q,\tilde \eta,\tilde p  )=
(\xi,q,\eta,p)
\end{displaymath}
defined by the Hamiltonian flow at time $t=1$ of $\chi_{N}$ as the Lie
series
\begin{equation}\label{eq:theCchin}
 \zeta = e^{\,L_{\chi_N}}\tilde \zeta :=\tilde \zeta + \{\tilde \zeta, \chi_N \}+
{1\over 2} \{ \{\tilde \zeta, \chi_N \}, \chi_N \}+\ldots  ~,
\end{equation}
where $L_{\chi_N} \equiv \{\cdot, \chi_N \}$ is the Poisson bracket operator 
and $\zeta,\tilde \zeta$ denote any of the variables $ \xi,q,\eta,p$ and
$\tilde \xi,\tilde q,\tilde \eta,\tilde p$ respectively.

\vskip 0.1 cm
\noindent
{\underline {\it Step 3:}} We compute the transformed Hamiltonian
\begin{equation}\label{eq:newham}
  {\cal K}^{(N)}_E = {\cal C}_{\chi_{N}}\, {\cal K}^{(N-1)}_E= e^{L_{\chi_N}}\, {\cal K}^{(N-1)}_E
\end{equation}
which, by construction, is in normal form up to order $N$, i.e. as in
Eq.~\eqref{normLC}.




In summary, for any given $N$  we obtain the transformation yielding the original variables $(\xi,q,\eta,p)$ in terms of the normalized 
variables $(\xi^{(N)},q^{(N)},\eta^{(N)},p^{(N)})$, via
  \begin{displaymath}
    \begin{aligned}
      (\xi,q,\eta,p) &= {\cal C}_N \,(\xi^{(N)},q^{(N)},\eta^{(N)},p^{(N)})\\
      & =  e^{L_{\chi_N}}\, \ldots  e^{L_{\chi_3}}\,
      (\xi^{(N)},q^{(N)},\eta^{(N)},p^{(N)})~.
    \end{aligned}
\end{displaymath}
The inverse transformation, i.e. the transformation yielding the normalized variables in terms of the original variables, is given by
  \begin{equation}\label{calcneq}
      (\xi^{(N)},q^{(N)},\eta^{(N)},p^{(N)}) = e^{-L_{\chi_3}}\, \ldots
      e^{-L_{\chi_N}}\,     (\xi,q,\eta,p)~.
\end{equation}
\vskip 0.4 cm
\noindent
{\bf Solutions of Hamilton equations of 
the normal form Hamiltonian.} Let us consider the solutions 
($\xi^{(N)}(\tau)$,$q^{(N)}(\tau)$,\\
$\eta^{(N)}(\tau)$,$p^{(N)}(\tau)$)
of the Hamilton's equations of \eqref{normLC},
\begin{displaymath}
{d \xi^{(N)}\over d\tau} = {\partial {\cal K}_E^{(N)}\over
\partial \eta^{(N)}}\ \ ,\ \ 
{d \eta^{(N)}\over d\tau} = -{\partial {\cal K}_E^{(N)}\over
\partial \xi^{(N)}}
\end{displaymath}
\begin{displaymath}
{d q^{(N)}\over d\tau} = {\partial {\cal K}_E^{(N)}\over
\partial p^{(N)}}\ \ ,\ \ 
{d p^{(N)}\over d\tau} = -{\partial {\cal K}_E^{(N)}\over
\partial q^{(N)}}
\end{displaymath}
on the zero-energy level of ${\cal K}_E^{(N)}$,  
transformed back to the 
variables $u,U$ using the transformations \eqref{eq:diagtrans} and 
\eqref{calcneq},
\begin{displaymath}
  \begin{aligned}
  (\xi,q,\eta,p)&=\left({\cal C}_N (\xi^{(N)},q^{(N)},\eta^{(N)},p^{(N)} \right)\\
(u_1,u_2,U_1,U_2)&=A (\xi,q,\eta,p) +(u_{*,1},u_{*,2},
U_{*,1},U_{*,2})~.
  \end{aligned}
\end{displaymath}
They project to solutions of the three-body problem of energy $E$ via
the Levi-Civita transformation. In fact, since the transformations
\eqref{eq:diagtrans} and \eqref{calcneq} are canonical, we only have
to prove that, for any initial condition
($\xi^{(N)}(0)$,$q^{(N)}(0)$,$\eta^{(N)}(0)$,$p^{(N)}(0)$) satisfying
\begin{displaymath}
{\cal K}^{(N)}_E \left(\xi^{(N)}(0),q^{(N)}(0),\eta^{(N)}(0),p^{(N)}(0) \right)=0
\end{displaymath}
transformed back to the variables $(u,U)$, we have 
\begin{displaymath}
{\cal K}_E \left(u_1(0),u_2(0),U_1(0),U_2(0) \right)=0 .
\end{displaymath}
This is a consequence of the 
definitions \eqref{eq:expham} and \eqref{normLC} 
of ${\cal K}^{(2)}_E$ and ${\cal K}^{(N)}_E$. In fact we have
\begin{displaymath}
  \begin{aligned}
0 &={\cal K}^{(N)}_E \left(\xi^{(N)}(0),q^{(N)}(0),\eta^{(N)}(0),p^{(N)}(0)\right) \\
& ={\cal K}^{(2)}_E
\left({\cal C}_N (\xi^{(N)}(0),q^{(N)}(0),\eta^{(N)}(0),p^{(N)}(0))\right) \\
&=
{\cal K}^{(2)}_E \left(\xi(0),q(0),\eta(0),p(0)\right) \\
&= 
{\cal K}_E \left(A (\xi(0),q(0),\eta(0),p(0)) +(u_{*,1},u_{*,2},
U_{*,1},U_{*,2}) \right)\\
&= {\cal K}_E \left(u_1(0),u_2(0),U_1(0),U_2(0) \right)  .
  \end{aligned}
  \end{displaymath}

\vskip 0.4 cm
\noindent
{\bf Computation of the Lyapunov orbits and their
stable and unstable manifolds.} Our aim is to use the normalizing 
transformations defined above to obtain analytic representations 
of the Lyapunov orbits as well as their local stable and unstable manifolds. 
To simplify the notation, from now on, we avoid the superscripts for
normalized variables.  

For any fixed value of $E>E_1$, we normalize the Hamiltonian
${\cal K}_E$ with a canonical transformation ${\cal C}_N$ (which
also depends parametrically on $E$) and obtain the normal form
Hamiltonian  ${\cal K}^{(N)}_E$. The Lyapunov orbit of energy $E$ 
is represented in the normalized variables by
\begin{equation}\label{llireg}
  \begin{aligned}
LL_1(E) &= \Big \{\xi, q,\eta,p:\\  &\xi,\eta=0 ,K_0+ i \omega q p+
\sum_{j=3}^{N}
K_{j}(0,q,0,p)=0 \Big \}~,
  \end{aligned}
\end{equation}
and then it is mapped back to the original variables using the inverse
of the canonical transformation ${\cal C}_N$ and all the
transformations needed to pass from the Cartesian variables
$(x,y,p_x,p_y)$ to the regularized variables $(u,U)$ and to the
variables $(\xi,q,\eta,p)$.
\vskip 0.4 cm
\noindent
The error in computing the set $LL_1(E)$ by formula (\ref{llireg}),
which neglects the remainder terms, depends on the norm of the
remainder in a small neighbourhood of the set $LL_1(E)$, which is well
represented by the quantity
\begin{equation}\label{normR}
  \begin{aligned}
  R^N & := \sup_{(\xi,q,\eta,p)\in LL_i(C)}
  \sum_{j \geq N+1} \hspace{-0.3cm}
  \sum_{\substack{ {k}, {l}\in {\Bbb N}^2:\\ \sum_n (k_n+l_n)=j}} \hspace{-0.4cm}
\text{\small $\norm{\alpha^{(N)}_{k_1,l_1,0,0}}$} \norm{q}^{k_1}\norm{p}^{l_1}\\
   & \leq \alpha_N \rho^{N+1}
  \end{aligned}
\end{equation}
where the coefficient $\alpha_N$ depends on $N$ and $\rho$
denotes the maximum amplitude in the variables $q,p$ on $LL_1(E)$.  In
principle, one wishes to use the largest possible value of $N$ which
is compatible with a reasonable CPU time and memory usage of modern
computers. However, as it is typical of normalizing transformations,
the coefficients $\alpha_N$ can increase with $N$ so that it may
happen that $\alpha_N \rho^{N+1}$ decreases only up to a certain value
of $N$, depending on $\rho$. The values of the coefficients $\alpha_N$
increase also as $E$ increases, since the values of $\lambda,\omega$,
which appear at the denominators of the generating functions $\chi_i$,
decrease. Moreover, since also $\rho$ increases as $E$ increases, we
have an increment of the error terms $\alpha_N \rho^{N+1}$ which
limits the validity of the method up to a certain value of $E$.

In this paper, we do not quantify a priori these errors, but we
set up a numerical method to estimate their effects.  In particular,
rather than constructing the stable and unstable sets
$W^s_1(E),W^u_1(E)$, we construct two surfaces, which we call the
inner and outer stable/unstable tubes, containing the sets
$W^s_1(E),W^u_1(E)$.

First, the stable and unstable manifolds $W^s_1(E),W^u_1(E)$ are
computed from the normalization implemented using strategy (a). In
this case, the sets are represented in the normalized variables by
\vspace{-0.15cm}
{\small
  \begin{equation*}
    \begin{aligned}
   W^s_1(E)  &= \Big \{\xi, q,\eta,p: xi=0 ,\ \ K_0+  i \omega q p+\sum_{j=3}^{N}
    K_{j}(0,q,\eta,p)=0\Big \}~,\\
   W^u_1 (E) &= \Big \{\xi,q,\eta,p: \eta=0 ,\ \ K_0+  i \omega q p+\sum_{j=3}^{N} K_{j}(\xi,q,0,p)=0\Big \}~,
    \end{aligned}
\end{equation*}}  
and then mapped back to the original Cartesian variables
$(x,y,p_x,p_y)$.  As a matter of fact, since the normalizing
transformation is valid only in a small neighbourhood of the set
$LL_1(E)$, we compute the following sections of the stable and
unstable manifolds:
{\small
  \begin{equation}\label{circles}
    \begin{aligned}
      &\Big \{\xi, q,\eta,p: \\
      &\xi=0 ,\eta=\eta_0,\ \ K_0+  i \omega q p +\sum_{j=3}^{N}  K_{j}(0,q,\eta_0,p)=0\Big \} \subseteq W^s_1(E)~,\\
      & \Big \{\xi,q,\eta,p: \\
      &\xi=\xi_0,\eta=0 , K_0+ i \omega q p + \sum_{j=3}^{N}
  K_{j}(\xi_0,q,0,p)=0\Big \} \subseteq W^u_1(E)~,
    \end{aligned}
\end{equation}}
with suitably small values of $\eta_0\ne 0,\xi_0\ne 0$. Then, the
tubes $ W^s_1(E),W^u_1(E)$ are constructed by computing numerically
the orbits with initial conditions in a grid of points of the sets
(\ref{circles}).
\vskip 0.4 cm
\noindent
We quantify the effect of the errors in the computation of the sets
$LL_1,W^s_1(E),W^u_1(E)$ as follows:

\vspace{0.3cm}
\noindent
$\quad (1) \,$ We consider a point in the set $LL_1$, obtained as explained
above. Then, we compute numerically the Hamilton equations of the
Levi-Civita Hamiltonian (\ref{eq:KofE}) with such initial condition
and we represent it in the Cartesian variables $(x(t),y(t))$. In the
case the initial condition is on the Lyapunov orbit, without any
error, the numerical integration provides a periodic orbit of period
$T$. Otherwise, the numerically integrated orbit does not exactly
closes after $T$, and the distance of $(x(T),y(T))$ from $(x(0),y(0))$
provides an estimate of the error in the computation of $LL_1$.

\vspace{0.3cm}
\noindent
$\quad(2) \,$ We consider for definiteness the stable manifold, and
its branch which extends on the right of the Lyapunov orbit towards
the singularity. The set $W^s_1(E)$ computed from the first of
equations (\ref{circles}) is affected by some error. As done in
\cite{GL18}, we can profit of a property of the tube manifolds to
construct two surfaces, the inner and outer stable tubes
$W^{s,in}_1(E)$, $W^{s,out}_1(E)$, that contain the true stable manifold
$W^s_1(E)$. As soon as the two surfaces $W^{s,in}_1(E),W^{s,out}_1(E)$
are very close, the numerical errors in the computation of $W^s_1(E)$
is small.

\noindent
The inner and outer stable tubes are defined as by computing
the two cycles {\small
  \begin{equation}\label{circlesinout}
    \begin{aligned}
      &\Big \{\xi, q,\eta,p: \\
      &\xi=\xi_0 ,\eta=\eta_0,\ \ K_0+i\omega  q p +
      \sum_{j=3}^{N} K_{j}(\xi_0,q,\eta_0,p)=0\Big \},\\
      &\Big \{\xi,q,\eta,p: \\
      &\xi=-\xi_0,\eta=\eta_0 ,\ \ K_0+ i\omega  q p+\sum_{j=3}^{N}
    K_{j}(-\xi_0,q,\eta_0,p)=0\Big \},
    \end{aligned}
  \end{equation}}
  with $0<\xi_0<<\eta_0$. The parameters $\xi_0,\eta_0$ are chosen so
that by computing the initial condition on the two cycles
(\ref{circlesinout}) forward in time, we observe that all the orbits
with initial conditions on one cycle approach the Lyapunov orbit and
then bounce back on the right, while all the orbits with initial
conditions on the other cycle approach the Lyapunov orbit and then
transit to the left. Since the stable manifold is a separatrix for the
motions which approach the Lyapunov orbit and then transit on the left
or bounce back to the right, by computing the backward evolution of
orbits with initial conditions in the cycles (\ref{circlesinout}) we
obtain two tubes containing the tube manifold $W^s_1(E)$.

In the next Sections we discuss the implementation of this theory and
we present numerical demonstrations of the method to the computation
of transit orbits.


\section{Efficiency of the normal form computations
for the Sun-Jupiter system}

\begin{figure*}
  \centering \includegraphics[width=2.0\columnwidth]{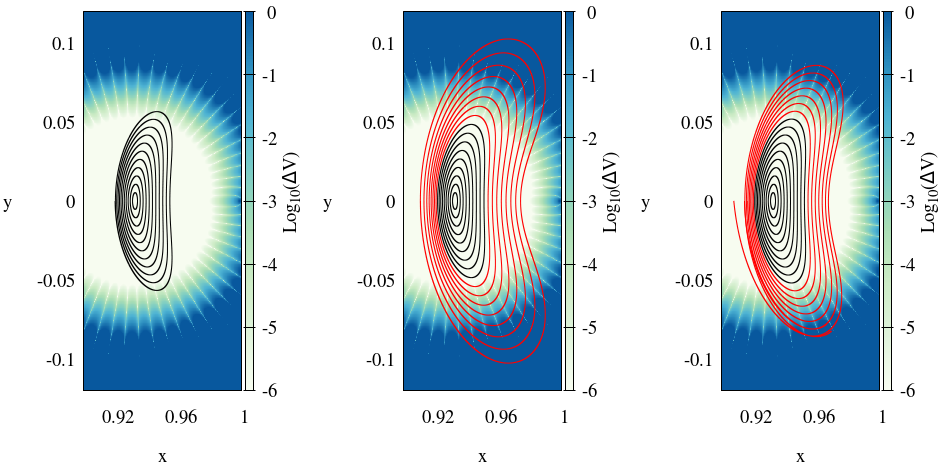}
  \caption{Lyapunov orbits of different amplitudes computed by
    normalizing the Hamiltonian in the Cartesian variables with the
    method (a) (left panel), method (b) (center panel) and method (c)
    (right panel). The initial conditions have been obtained from the
    normalized Cartesian Hamiltonian, while the orbits have been
    numerically computed by integrating the regularized Hamiltonian
    ${\cal K}_E$ in Eq.~\eqref{eq:KofE}. On the
      background we plot with a color scale the logarithm of 
      $\norm{\Delta V_N}$ defined in (\ref{deltaVn}). 
The orbits in black are those for which the energy error 
is in agreement with the normal form (see eq. (\ref{Equad}) and 
Fig. \ref{fig:departure}).}
  \label{fig:cartesian}
\end{figure*}

\begin{figure*}
  \includegraphics[width=2.\columnwidth]{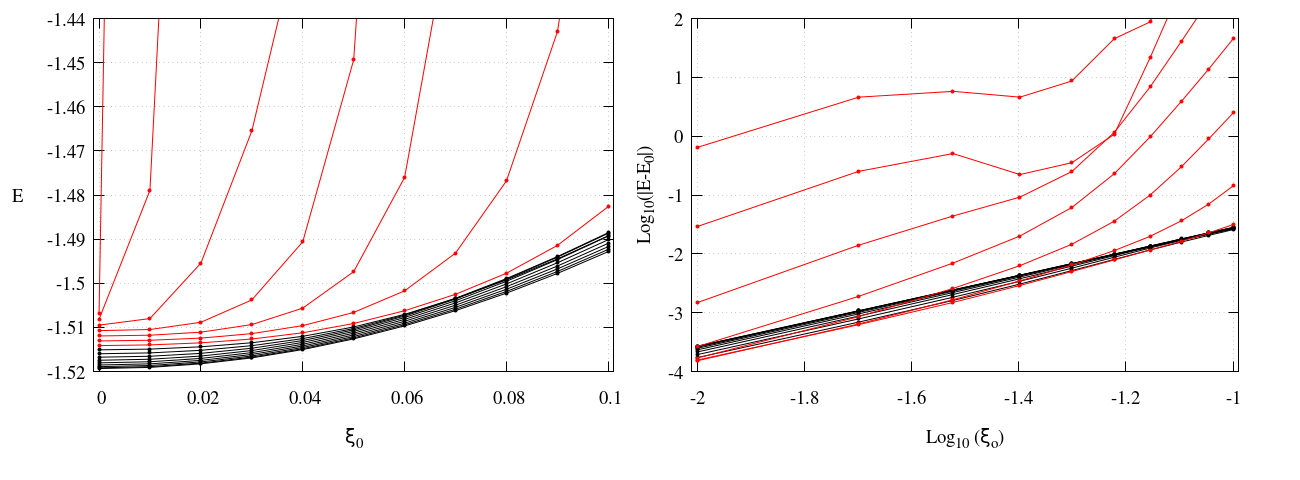}
  \caption{The curves in both panels represent  the quantity $E$ defined in 
eq. (\ref{Equad}) computed for $q_0,p_0$ corresponding to a selected point in each Lyapunov
orbit of  Fig. \ref{fig:cartesian}, while $\eta=0$ and $\xi\in [0,0.1]$. 
The black and red curves correspond to the Lyapunov orbits represented in
black and red respectively. We appreciate that only for the black curves
the behaviour of $E$ is in agreement with eq. (\ref{Equad}).  }
  \label{fig:departure}
\end{figure*}

Aiming to investigate the limits of the normalizations implemented
with the Cartesian variables, we represent the family of Lyapunov
orbits $LL_1(E)$ at different values of the energy, computed from the
normalization of the Hamiltonian expressed with these variables with
the three different normalizing strategies (a), (b), (c), see
Fig.~\ref{fig:cartesian}. The starting point of any
  Hamiltonian normalization in Cartesian variables includes the
  computation of the Taylor expansion $V_N$ at order $N$ of the
  singular potential energy
\begin{displaymath}
V= - {\mu\over
\sqrt{(x-1+\mu)^2+y^2}}
\end{displaymath}
appearing in Hamiltonian (\ref{hamcartesiana}). As already remarked, 
for $\mu=\mu_J$, the function $V$ has complex singularities located at 
$x=x_{L_1},y= \pm i \sigma \sim 0.06 \pm i 0.066...$. For the values of the 
energy $E$ that we consider, we need to perform normalizations for 
$N\in [20,30]$. In this range of $N$ the complex singularity affects the 
Taylor expansions for $y> 0.066$. For example, by denoting with 
\begin{equation}\label{deltaVn}
\Delta V_N(x,y)=
\norm{V_N(x,y)- V(x,y)}  ,
\end{equation}
an easy computation shows
\begin{center}
\begin{tabular}{l c c c}\label{deltaV}
  $ $ &   $ $ &   $ $ &   $ $ \\
  \hline
   $N$  &  $\Delta V_N \text{\small $(x_{L_1},0.066$)}$   &   $\Delta V_N \text{\small $(x_{L_1},0.08$)}$ & $\Delta V_N \text{\small $(x_{L_1},0.1$)}$ \\
   \hline
20 &  $10^{-3}$ &  $5.5\ 10^{-2}$      &    $5.7$ \\
30  &    $7.4\ 10^{-4}$ & $0.28$ &  $270$ \\
\hline
  $ $ &   $ $ &   $ $ &   $ $ 
 \end{tabular}
\end{center}
\vskip 0.4 cm In order to check how this singularity
  affects the validity of the normal form expansions, in
  Fig.~\ref{fig:cartesian} we show the Lyapunov orbits corresponding
  to $E \in [-1.5193,-1.502]$ obtained from the normalization of the
  Cartesian Hamiltonian with the three strategies (a), (b), (c) for
  $N=30$. The orbits in Fig.~\ref{fig:cartesian} are obtained by
numerically integrating an initial condition of the approximate
Lyapunov orbit obtained from Eq.~\eqref{lli}, without applying
differential corrections\footnote{Since our aim here is to compare the
  limits of the different methods of normalization, we prefer not to
  reduce it with a differential correction of the initial
  conditions.}.  Due to the abundance of bibliography explaining this
type of computations (see \cite{jorbamas99},~\cite{giorgilli12} and
references therein), we limit ourselves to present only the results.
In the background we plot in color scale the value of
  the function $\log_{10}\norm{\Delta V_N}$. With evidence, the
  largest reliable Lyapunov orbit obtained with the Cartesian
  variables strategy (a) (Fig.~\ref{fig:cartesian}, left panel) is
  within the domain of the plane $x,y$ where the error function
  $\norm{\Delta V_N(x,y)}$ is small. Instead, we observe that for the
  largest Lyapunov orbit obtained with the Cartesian
  variables-strategy (b) (Fig.~\ref{fig:cartesian}, center panel), the
  amplitude is well beyond the limit of the complex singularity of
  $V$. As a matter of fact, the effect of the error due to the
  difference $\norm{\Delta V_N}$ appears when we consider values of
  the hyperbolic variables $\xi,\eta\ne 0$. From the specific
  expression of the normal form $(b)$ (Eq.~(\ref{normLCb})),
  for any given $(q_0,p_0)$ we have:
\begin{equation}\label{Equad}
E= E_0(q_0,p_0)+ \xi^2 A(q_0,p_0)+.....
\end{equation}
and therefore we expect a quadratic growth of $E-E_0(q_0,p_0)$ with
$\xi$. This behavior is confirmed only for the Lyapunov orbits in
black in Fig.~\ref{fig:cartesian}, center panel.  Precisely, by
choosing a sample point for each Lyapunov orbit of
Fig.~\ref{fig:cartesian}, given by $q_0,p_0,\xi=0,\eta=0$ in the
normalized variables, in Fig.~\ref{fig:departure} we show the values
of the energy $E$ computed for $\eta=0$ and $\xi\in [0,1/10]$. Only
for the Lyapunov orbits represented in black we have an agreement
between the curves of Fig.~\ref{fig:departure} and the law
(\ref{Equad}). The same behaviour is found for the normalization
method (c) (Fig.~\ref{fig:cartesian}, right panel).  We conclude that
the limit of the normalizations implemented with the Cartesian
variables is given by $\norm{1-\mu-x_{L_1}}$.

\begin{figure}
  \centering
  \includegraphics[width=\columnwidth]{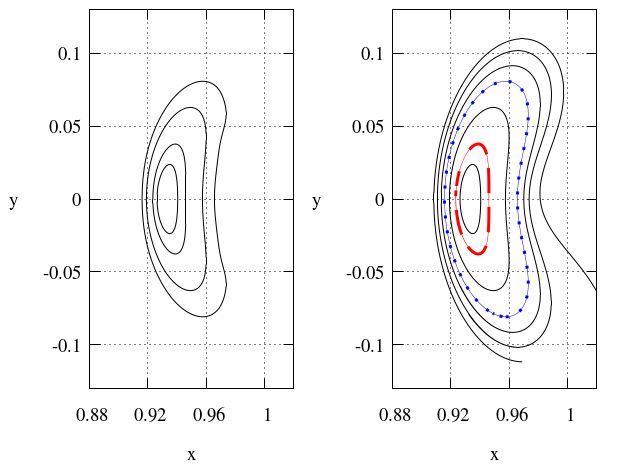}
  \caption{Lyapunov orbits of different amplitudes computed by
    normalizing the Levi-Civita Hamiltonian with the method (a) (left
    panel) and method (c) (right panel). The initial conditions have
    been obtained from the normalized Levi-Civita Hamiltonian, while
    the orbits have been numerically computed by integrating the
    non-normalized Hamiltonian ${\cal K}_E$.}
  \label{fig:regularized}
\end{figure}

\begin{table}
 \caption{Norm of the remainder $R^N$ evaluated on the Lyapunov orbit
   for $E_a=-1.516$ (left columns) and for $E_b=-1.509$ (right
   columns), according to Eq.~\eqref{normR}, produced with expansions
   limited to order ${\cal N}_p = 27$ and normalization order $N =
   3,...,24$.}
 \label{tab:normdata0}
 \begin{center}
 \begin{tabular}{llll}
   \hline \multicolumn{2}{c}{Red orbit Fig.~\ref{fig:regularized}} & \multicolumn{2}{c}{Blue orbit Fig.~\ref{fig:regularized}} \\
   $N$ & $ R^N$ & \phantom{X} $N$ &
   $R^N$\\ \hline $3$ & $1.61835
   \,\snot[-5]$ & \phantom{X} $3$ & $7.28121 \,\snot[-4]$ \\ $4$ &
   $2.35741 \,\snot[-6]$ & \phantom{X} $4$ & $3.53145 \,\snot[-4]$
   \\ $5$ & $4.90253 \,\snot[-7]$ & \phantom{X} $5$ & $2.61175
   \,\snot[-4]$ \\ $6$ & $1.00808 \,\snot[-7]$ & \phantom{X} $6$ &
   $1.98361 \,\snot[-4]$ \\ $7$ & $2.34993 \,\snot[-8]$ & \phantom{X}
   $7$ & $1.45230 \,\snot[-4]$ \\ $8$ & $4.72191 \,\snot[-9]$ &
   \phantom{X} $8$ & $1.04454 \,\snot[-4]$ \\ $9$ & $2.13184
   \,\snot[-9]$ & \phantom{X} $9$ & $8.94943 \,\snot[-5]$ \\ $10$ &
   $4.13041 \,\snot[-10]$ & \phantom{X} $10$ & $6.06359 \,\snot[-5]$
   \\ $11$ & $1.51257 \,\snot[-10]$ & \phantom{X} $11$ & $4.88443
   \,\snot[-5]$ \\ $12$ & $3.59044 \,\snot[-11]$ & \phantom{X} $12$ &
   $3.31315 \,\snot[-5]$ \\ $13$ & $8.67395 \,\snot[-12]$ &
   \phantom{X} $13$ & $2.28439 \,\snot[-5]$ \\ $14$ & $1.81004
   \,\snot[-12]$ & \phantom{X} $14$ & $1.53451 \,\snot[-5]$ \\ $15$ &
   $5.38462 \,\snot[-13]$ & \phantom{X} $15$ & $1.14096 \,\snot[-5]$
   \\ $16$ & $9.09849 \,\snot[-14]$ & \phantom{X} $16$ & $7.54539
   \,\snot[-6]$ \\ $17$ & $2.32918 \,\snot[-14]$ & \phantom{X} $17$ &
   $5.71248 \,\snot[-6]$ \\ $18$ & $5.61067 \,\snot[-15]$ &
   \phantom{X} $18$ & $4.31952 \,\snot[-6]$ \\ $19$ & $2.70410
   \,\snot[-15]$ & \phantom{X} $19$ & $3.64058 \,\snot[-6]$ \\ $20$ &
   $3.35487 \,\snot[-16]$ & \phantom{X} $20$ & $2.42248 \,\snot[-6]$
   \\ $21$ & $1.67395 \,\snot[-16]$ & \phantom{X} $21$ & $2.22411
   \,\snot[-6]$ \\ $22$ & $4.91488 \,\snot[-17]$ & \phantom{X} $22$ &
   $1.63076 \,\snot[-6]$ \\ $23$ & $1.75187 \,\snot[-17]$ &
   \phantom{X} $23$ & $1.31428 \,\snot[-6]$ \\ $24$ & $5.80105
   \,\snot[-18]$ & \phantom{X} $24$ & $9.80338 \,\snot[-7]$ \\ \hline
 \end{tabular}
 \end{center}
\end{table}

We therefore proceed to check that the normalization of the
Hamiltonian expressed with the Levi-Civita variables overcomes this
limit.  First, in Fig.~\ref{fig:regularized} we show the Lyapunov
orbits computed using method (a) in the left panel and method (c) in
the right panel. The initial conditions used for each case have been
obtained from the normalized Levi-Civita Hamiltonian ${\cal
  K}^{(N)}_E$ in Eq.~\eqref{normLC} at the specific energy value, and
they were numerically integrated with the non-normalized Hamiltonian
${\cal K}_{E}$~\eqref{eq:KofE}. Thanks to the implementation of 
the normalization method in the regularized Levi-Civita Hamiltonian,  
 with both methods we compute 
Lyapunov orbits of libration amplitudes larger than the threshold
value $0.066...$ of the Cartesian normalizations. Then, the threshold value
of the largest Lyapunov orbit depends also on the normalization method. 
In fact, with the method (c)
we reach a larger value for the amplitude with
respect to the method (a). The limiting factor for strategy (a) is
clearly the large number of terms to eliminate to render the
Hamiltonian in normal form. Since for the computation of the stable
and unstable manifolds we are forced to use method (a), we have an
indication of the improvement introduced by the Levi-Civita variables,
which allow us to reach a libration amplitude of about $1.4
\ \norm{1-\mu-x_{L_1}}$. 

With the sake of explaining the differences between these techniques,
we provide a few specific examples of the normal forms computations.
We consider here the computation of Lyapunov orbits for two specific
values of the energy, the first corresponding to a small amplitude
Lyapunov orbit ($E = E_a :=-1.516$) and the second corresponding to a
moderate amplitude Lyapunov orbit ($E= E_b := -1.509$).  For
reference, $LL_1(E_a)$ is represented with a dashed red curve in
Fig.~\ref{fig:regularized}, while $LL_1(E_b)$ corresponds to the
dotted blue orbit. We expect that small libration orbits imply small
errors in the computation.  This is confirmed by the computation of
the norm of the remainders $R^N$ (evaluated according to
Eq.~\eqref{normR}) of such orbits, represented in
Table~\ref{tab:normdata0}, for $N\in \{3,\ldots ,24\}$, for the
normalizing strategy (c).

Additionally, in Tables~\ref{tab:normdata1} and~\ref{tab:normdata2},
we compare the normal forms obtained with the three strategies for the
first two steps of the normalization, giving ${K_3}$ and ${K_4}$
respectively, for $LL_1(E_b)$. Each entry of the tables refers to a
specific monomial appearing in the normal form, for which it is given
the integers $k_1$, $l_1$, $k_2$, $l_2$, indicated in which normal
form it is present (out of the three strategies), and the associated
coefficient $\alpha$.

\begin{table}
 \caption{Normal form decomposition of ${K_3}$ for the example
   $LL_1(E_a)$}
 \label{tab:normdata1}
 \begin{center}
 \begin{tabular}{llll|lll|c}
   \hline
   \multicolumn{4}{c}{monomial} & \multicolumn{3}{c}{strategy} &   \phantom{$\alpha$} \\
   {\scriptsize $k_1$} & {\scriptsize $l_1$} & {\scriptsize $k_2$} & {\scriptsize $l_2$} & {\scriptsize (a)} & {\scriptsize (b)} & {\scriptsize (c)} & $\alpha^{(3)}_{k_1,l_1,k_2,l_2}$ \\
   \hline
   0 & 0 & 0 & 3 & \textcolor{white}{x} & \textcolor{black}{x} & \textcolor{black}{x} & \multicolumn{1}{l}{$-0.239846$} \\
   0 & 1 & 0 & 2 & \textcolor{white}{x} & \textcolor{black}{x} & \textcolor{black}{x} & $0.1371133 + 0.0334760 \,\mathrm{i}\,$ \\
   0 & 3 & 0 & 0 & \textcolor{black}{x} & \textcolor{white}{x} & \textcolor{black}{x} & \multicolumn{1}{l}{$0.144918$} \\
   1 & 0 & 0 & 2 & \textcolor{white}{x} & \textcolor{black}{x} & \textcolor{black}{x} & $0.0334760 + 0.1371133 \, \mathrm{i}\,$ \\
   1 & 2 & 0 & 0 & \textcolor{black}{x} & \textcolor{white}{x} & \textcolor{black}{x} & \multicolumn{1}{l}{$-0.164176\, \mathrm{i}$}\\
   2 & 1 & 0 & 0 & \textcolor{black}{x} & \textcolor{white}{x} & \textcolor{black}{x} & \multicolumn{1}{l}{$0.164176$} \\
   3 & 0 & 0 & 0 & \textcolor{black}{x} & \textcolor{white}{x} & \textcolor{black}{x} & \multicolumn{1}{l}{$-0.144918\, \mathrm{i}$} \\
   0 & 0 & 1 & 2 & \textcolor{white}{x} & \textcolor{black}{x} & \textcolor{black}{x} & \multicolumn{1}{l}{$0.642141$} \\
   0 & 1 & 1 & 1 & \textcolor{black}{x} & \textcolor{black}{x} & \textcolor{black}{x} & \multicolumn{1}{l}{$-0.284350$} \\
   1 & 0 & 1 & 1 & \textcolor{black}{x} & \textcolor{black}{x} & \textcolor{black}{x} & \multicolumn{1}{l}{$-0.284350\, \mathrm{i}$} \\
   0 & 0 & 2 & 1 & \textcolor{white}{x} & \textcolor{black}{x} & \textcolor{black}{x} & \multicolumn{1}{l}{$-0.642141$} \\
   0 & 1 & 2 & 0 & \textcolor{white}{x} & \textcolor{black}{x} & \textcolor{black}{x} & $0.1371133 - 0.0334760 \, \mathrm{i}\,$ \\
   1 & 0 & 2 & 0 & \textcolor{white}{x} & \textcolor{black}{x} & \textcolor{black}{x} & $-0.0334760 + 0.1371133\, \mathrm{i}\,$ \\
   0 & 0 & 3 & 0 & \textcolor{white}{x} & \textcolor{black}{x} & \textcolor{black}{x} & \multicolumn{1}{l}{$0.239846$} \\
   \hline
   \multicolumn{4}{c}{Total of terms} & 6 & 10 & 14 & \phantom{lalala} \\
   \hline
 \end{tabular}
 \end{center}
\end{table}

\begin{table}[t]
 \caption{Normal form decomposition of ${K_4}$ for the example
   $LL_1(E_a)$}
 \label{tab:normdata2}
 \begin{tabular}{llll|lll|cl}
   \hline
   \multicolumn{4}{c}{monomial} & \multicolumn{3}{c}{strategy} & \phantom{$\alpha$} & \phantom{A} \\
               {\scriptsize $k_1$} & {\scriptsize $l_1$} & {\scriptsize $k_2$} & {\scriptsize $l_2$} &  {\scriptsize (a)} & {\scriptsize (b)}
               & {\scriptsize (c)} & $\alpha^{(4)}_{k_1,l_1,k_2,l_2}$ & \phantom{x}\\
   \hline
   0 & 0 & 0 & 4 & \textcolor{white}{x} & \textcolor{black}{x} & \textcolor{black}{x} & \multicolumn{1}{l}{\small $-0.478035$} & \textcolor{white}{$^{\dagger}$}\\
   0 & 1 & 0 & 3 & \textcolor{white}{x} & \textcolor{black}{x} & \textcolor{black}{x} & {\small $0.732257 + 1.135126 \, \mathrm{i}$} & \textcolor{white}{$^{\dagger}$}\\
   0 & 2 & 0 & 2 & \textcolor{white}{x} & \textcolor{black}{x} & \textcolor{black}{x} & {\small $-0.342835 - 1.071743 \, \mathrm{i}$} & \textcolor{black}{$^{\dagger}$}\\
   0 & 4 & 0 & 0 & \textcolor{black}{x} & \textcolor{white}{x} & \textcolor{black}{x} & \multicolumn{1}{l}{\small $0.0583902$} & \textcolor{white}{$^{\dagger}$}\\
   1 & 0 & 0 & 3 & \textcolor{white}{x} & \textcolor{black}{x} & \textcolor{black}{x} & {\small $1.135126 + 0.732257 \, \mathrm{i}$} & \textcolor{white}{$^{\dagger}$}\\
   1 & 1 & 0 & 2 & \textcolor{white}{x} & \textcolor{black}{x} & \textcolor{black}{x} & \multicolumn{1}{l}{\small $-2.41906\, \mathrm{i}$} & \textcolor{black}{$^{\dagger}$}\\
   1 & 3 & 0 & 0 & \textcolor{black}{x} & \textcolor{white}{x} & \textcolor{black}{x} & \multicolumn{1}{l}{\small $-1.81534\, \mathrm{i}$} & \textcolor{white}{$^{\dagger}$}\\
   2 & 0 & 0 & 2 & \textcolor{white}{x} & \textcolor{black}{x} & \textcolor{black}{x} & {\small $0.342835 - 1.071743 \, \mathrm{i}$} & \textcolor{black}{$^{\dagger}$}\\
   2 & 2 & 0 & 0 & \textcolor{black}{x} & \textcolor{black}{x} & \textcolor{black}{x} & \multicolumn{1}{l}{\small $-3.37708$} & \textcolor{black}{$^{\dagger}$}\\
   3 & 1 & 0 & 0 & \textcolor{black}{x} & \textcolor{white}{x} & \textcolor{black}{x} & \multicolumn{1}{l}{\small $1.81534\,\mathrm{i}$} & \textcolor{white}{$^{\dagger}$}\\
   4 & 0 & 0 & 0 & \textcolor{black}{x} & \textcolor{white}{x} & \textcolor{black}{x} & \multicolumn{1}{l}{\small $0.0583902$} &\textcolor{white}{$^{\dagger}$}\\
   0 & 0 & 1 & 3 & \textcolor{white}{x} & \textcolor{black}{x} & \textcolor{black}{x} & \multicolumn{1}{l}{\small $2.23041$} & \textcolor{white}{$^{\dagger}$}\\
   0 & 1 & 1 & 2 & \textcolor{white}{x} & \textcolor{black}{x} & \textcolor{black}{x} & {\small $-2.86948 - 1.59924 \, \mathrm{i}$} & \textcolor{white}{$^{\dagger}$}\\
   0 & 2 & 1 & 1 & \textcolor{black}{x} & \textcolor{black}{x} & \textcolor{black}{x} & \multicolumn{1}{l}{\small $2.31725$} & \textcolor{black}{$^{\star}$}\\
   1 & 0 & 1 & 2 & \textcolor{white}{x} & \textcolor{black}{x} & \textcolor{black}{x} & {\small $-1.59924 - 2.86948\, \mathrm{i}$} & \textcolor{white}{$^{\dagger}$}\\
   1 & 1 & 1 & 1 & \textcolor{black}{x} & \textcolor{black}{x} & \textcolor{black}{x} & \multicolumn{1}{l}{\small $2.55115 \,\mathrm{i}$} & \textcolor{black}{$^{\star}$}\\
   2 & 0 & 1 & 1 & \textcolor{black}{x} & \textcolor{black}{x} & \textcolor{black}{x} & \multicolumn{1}{l}{\small $-2.31725$} & \textcolor{black}{$^{\star}$}\\
   0 & 0 & 2 & 2 & \textcolor{black}{x} & \textcolor{black}{x} & \textcolor{black}{x} & \multicolumn{1}{l}{\small $3.75798$} & \textcolor{black}{$^{\bullet}$}\\
   0 & 1 & 2 & 1 & \textcolor{white}{x} & \textcolor{black}{x} & \textcolor{black}{x} & {\small $- 2.86948 -1.59924\, \mathrm{i}$} & \textcolor{white}{$^{\dagger}$}\\
   0 & 2 & 2 & 0 & \textcolor{white}{x} & \textcolor{black}{x} & \textcolor{black}{x} & {\small $-0.342835 + 1.071743\, \mathrm{i}$}& \textcolor{black}{$^{\dagger}$}\\
   1 & 0 & 2 & 1 & \textcolor{white}{x} & \textcolor{black}{x} & \textcolor{black}{x} & {\small $-1.59924 + 2.86948\, \mathrm{i}$} & \textcolor{white}{$^{\dagger}$}\\
   1 & 1 & 2 & 0 & \textcolor{white}{x} & \textcolor{black}{x} & \textcolor{black}{x} & \multicolumn{1}{l}{small $-2.41906\, \mathrm{i}$} & \textcolor{black}{$^{\dagger}$}\\
   2 & 0 & 2 & 0 & \textcolor{white}{x} & \textcolor{black}{x} & \textcolor{black}{x} & {\small $0.342835 + 1.071743\, \mathrm{i}$} & \textcolor{black}{$^{\dagger}$}\\
   0 & 0 & 3 & 1 & \textcolor{white}{x} & \textcolor{black}{x} & \textcolor{black}{x} & \multicolumn{1}{l}{\small $2.23041$} & \textcolor{white}{$^{\dagger}$}\\
   0 & 1 & 3 & 0 & \textcolor{white}{x} & \textcolor{black}{x} & \textcolor{black}{x} & {\small $-0.732257 + 1.135126 \, \mathrm{i}$} & \textcolor{white}{$^{\dagger}$}\\
   1 & 0 & 3 & 0 & \textcolor{white}{x} & \textcolor{black}{x} & \textcolor{black}{x} & {\small $1.135126 - 0.732257\, \mathrm{i}$} & \textcolor{white}{$^{\dagger}$}\\
   0 & 0 & 4 & 0 & \textcolor{white}{x} & \textcolor{black}{x} & \textcolor{black}{x} & \multicolumn{1}{l}{\small $-0.478035$} & \textcolor{white}{$^{\dagger}$}\\
   \hline
   \multicolumn{4}{c}{Total of terms} & 9 & 23 & 27 & \multicolumn{2}{c}{\phantom{lalala}} \\
   \hline
 \end{tabular}
\end{table}

From Table~\ref{tab:normdata1}, we can see that the less
demanding normalization schemes (b) and (c) provide normal forms
with many more terms than the strategy (a). At this step of
normalization, there are no differences in the coefficients, since the
three strategies are applied to the same initial expansion (the terms
$K_0$ and $K_2$ are common to the three). When performing the
following step, since now the schemes are applied also on $K_3$, the
values of the coefficients start differing from strategy to
strategy. The coefficients in Table~\ref{tab:normdata1} refer to the
normalization by strategy (c), and we denote with different symbols
the monomials whose coefficient has a different value for the strategy
(a) ($\bullet$), the strategy (b) ($\dagger$) or both ($\star$). As we
proceed with the normalizations, the resulting $K_j$ are different
from each other. We also notice the increase in the amount of terms
from step $3$ to $4$. This effect is evident also in
Table~\ref{tab:normdata4}, which summarizes the information related to
the normal form computation (following strategy (c)) for the set of
Lyapunov orbits appearing in Fig.~\ref{fig:regularized} (right panel).
The information provided corresponds to: the values of $E$,
the order of the initial polynomial expansion $N_p$\footnote{As rule
  of thumb, we use an initial polynomial expansion at least three
  orders larger than the normalization order used, i.e. $N_p \geq N +
  3$.}, the normalization order $N$, the amount of terms of the normal
form ${\cal K}_E^{(N)}$ and of the remainder ${\cal R}_E^{(N)}$ at
order $N$, and the maximum distance that the orbit reaches with
respect to the position of $L_1$.

\begin{table}
 \caption{Normalization information for the orbits presented in
   Fig.~\ref{fig:regularized}.}
 \label{tab:normdata4}
 \begin{center}
 \begin{tabular}{lllllll}
  \hline $E$ & $N_p$ & $N$ & term ${\cal K}^{(N)}_E$ & terms
  ${\cal R}^{(N)}_E$ & Max ampl \\
  \hline $-1.518$ &  13 & 10 & $1658$ & $2056$ & $2.38 \snot[-2]$ \\
  $-1.516$  & 13 & 10 & $1658$ & $2056$ & $3.83 \snot[-2]$ \\
  $-1.512$ & 19 & 16 & $6967$ & $5335$ & $6.53 \snot[-2]$ \\
  $-1.509$ & 21 & 18 & $10200$ & $6924$ & $8.54 \snot[-2]$ \\
  $-1.507$ & 25 & 22 & $19869$ & $10990$ & $9.73 \snot[-2]$ \\
  $-1.505$ & 33 & 30 & $57843$ & $23346$ & $1.08 \snot[-1]$ \\
  $-1.504$ & 33 & 30 & $57843$ & $23346$ & $1.17 \snot[-1]$ \\
  \hline
 \end{tabular}
 \end{center}
\end{table}

As for increasing energies the periodic orbits have larger amplitudes,
it is necessary to use larger normalizations orders (see
Table~\ref{tab:normdata0}). Notice, for instance, that a variation of
$\sim 1$\% in the energy (from $E=-1.518$ to $E=-1.505$) requires to
use a normalization order 3 times larger, and the associated normal
form is 35 times longer. This exponential growth makes hardly
tractable the computation of these normal forms for larger values of
the energy.


\section{Numerical computation of the stable tubes manifolds}

\begin{figure}
  \centering \includegraphics[width=\columnwidth]{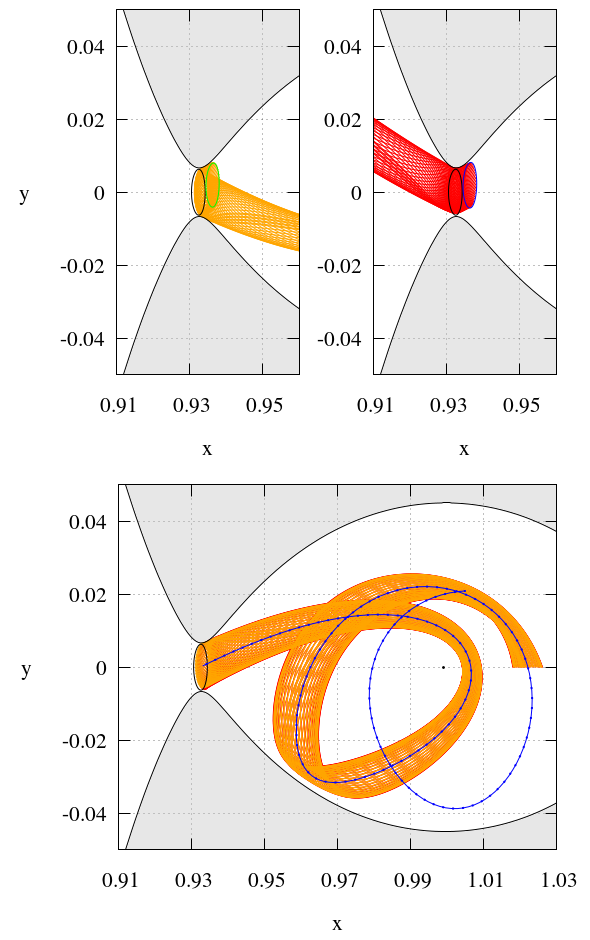}
  \caption{Representation of the inner and outer stable tubes for $ E
    =-1.5193$. {\bf Top panels:} the Lyapunov orbit $LL_1(E=-1.5193)$
    (black curve), initial conditions of the outer tube (green curve)
    and the inner tube (blue curve), orbits of the outer tube (orange)
    and the inner tube (red) integrated forward in time.  {\bf Bottom
      panel:} Outer (orange) and inner (red) tubes integrated backward
    in time, almost overlapped (see text for discussion), and the
    invariant stable manifold emanating from $LL_1$ (blue curve,
    appearing also in the top left panel of Fig.~\ref{fig:tubes}). The
    shaded area represents the realm of forbidden motions for this
    value of the energy.}
  \label{fig:tubes258}
\end{figure}

\begin{figure}
  \centering \includegraphics[width=\columnwidth]{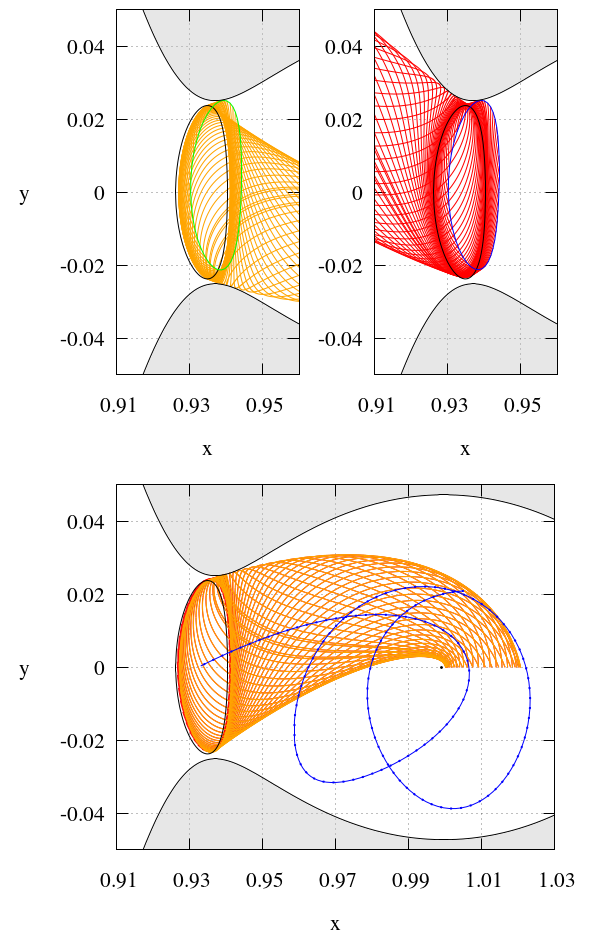}
    \caption{As Fig.~\ref{fig:tubes258}, for $E = -1.5183$.}
  \label{fig:tubes255}
\end{figure}

\begin{figure}[t]
  \centering \includegraphics[width=\columnwidth]{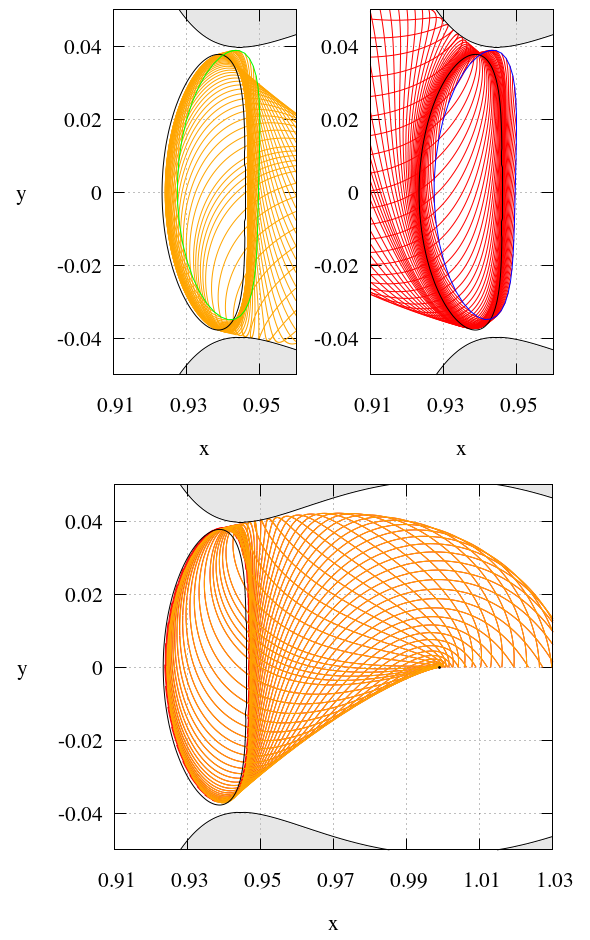}
  \caption{As Fig.~\ref{fig:tubes258}, for $E = -1.5167$.}
  \label{fig:tubes25}
\end{figure}

\begin{figure}[t]
  \centering \includegraphics[width=0.85\columnwidth]{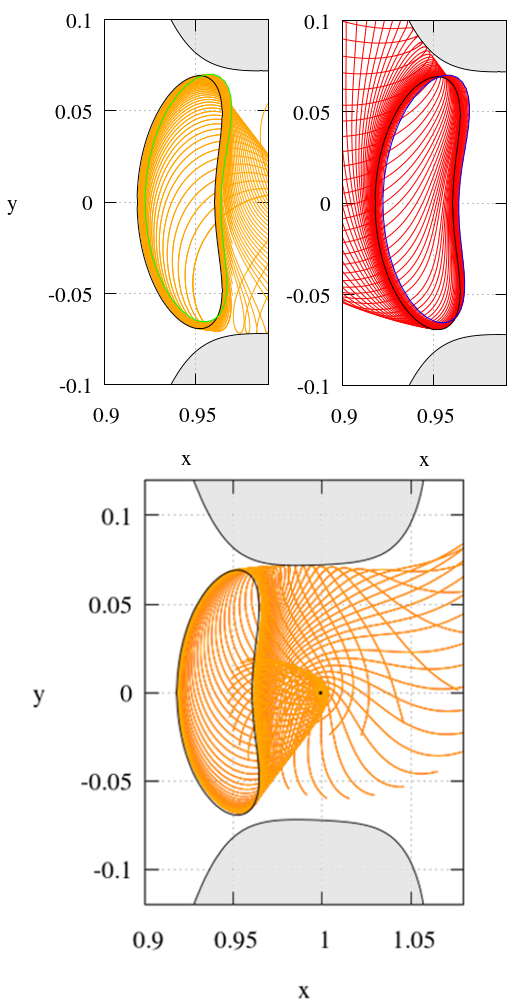}
  \caption{As Fig.~\ref{fig:tubes258}, for $E = -1.5113$.}
  \label{fig:tubes23}
\end{figure}

\begin{figure*}
  \centering \includegraphics[width=1.95\columnwidth]{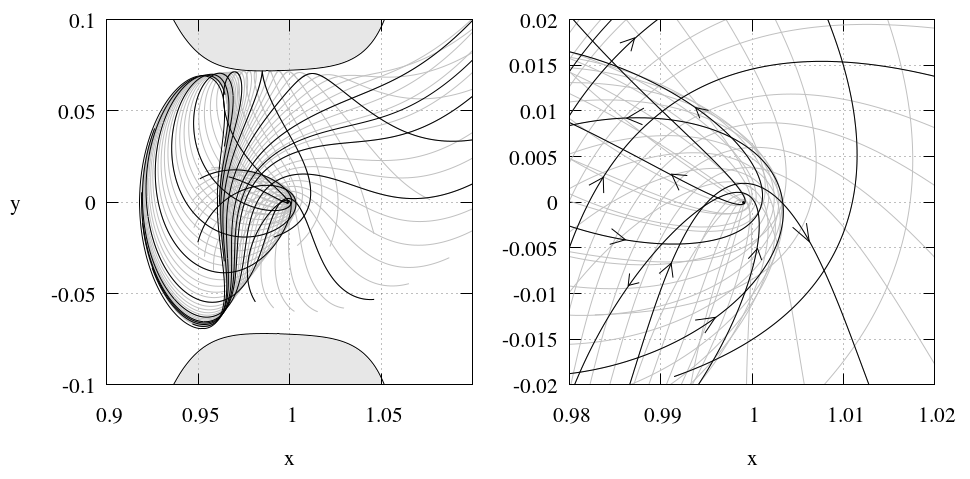}
  \caption{{\bf Left panel:} Same as Fig.~\ref{fig:tubes23} where some
    orbits are highlighted in bold black curves.  {\bf Right panel:}
    Zoomed-in of the left panel, in the vicinity of the primary
    $P_2$. The arrows indicate the sense of motion backward in
    time. Some of the orbits circulate clockwise, while others do it
    counter-clockwise.}
  \label{fig:zooms}
\end{figure*} 

In Fig.~\ref{fig:tubes258}, \ref{fig:tubes255}, \ref{fig:tubes25} and
\ref{fig:tubes23}, we represent the inner and outer stable tubes
computed using the normalized Levi-Civita Hamiltonian as explained in
Section 2 (see Eq.~\eqref{circlesinout} and related discussion), for
$E = -1.5193$, $-1.5183$, $-1.5167$, $-1.5113$ respectively.
For each value of $E$, the inner/outer tubes are
  obtained first by considering $\xi_0=\pm 10^{-4}<< \eta_0= 10^{-2}$,
  and by numerically computing a set of points $(q_j,p_j)$
  ($j=0,\ldots ,30$) sampling the level set
\begin{displaymath}
{\cal K}^{(N)}_E(\xi_0,q,\eta_0,p)=0  .
\end{displaymath}
Then, we compute the image of all the points
$(\xi_0,q_j,\eta_0,p_j)$ in the Levi-Civita variables $u,U$. Finally,
the tubes are obtained by numerically integrating the Hamilton's
equation of the Levi-CivitaHamiltonian ${\cal K}_E(u,U)$ for these
initial conditions.

On the top panels of each figure we represent separately the orbits with
initial conditions in the inner tube (blue line) and the orbits with
initial conditions in the outer tube (green line). When the initial
conditions on the outer stable tube are integrated forward in time, as
it happens in the top panels, initially they approach the Lyapunov
orbit (black curve), but then the orbits of the outer stable tube
(orange orbits) bounce back, while the orbits of the inner stable tube
(red orbits) transit to the left of the Lyapunov orbit. This
mechanism allows us to identify one tube from the other. When
integrated backward in time (bottom panels of Fig.~\ref{fig:tubes258},
\ref{fig:tubes255}, \ref{fig:tubes25}, \ref{fig:tubes23}), the orbits
of both inner and outer stable tubes almost overlap, indicating the
location of the invariant stable manifold.

Figures~\ref{fig:tubes258} and \ref{fig:tubes255} show the traditional
behavior for the stable manifold, as described at (ii) in Section
1. When integrated backward in time, both tubes fold around Jupiter
clockwise, following the sense of the collapsed invariant manifold
emanating from $LL_1(E_1)$ (blue line in the bottom panels),
$W^s_1(E_{1})$. The smaller the value of the energy is, the better
$W^s_1(E_{1})$ represent the behavior of the tube manifold, as it is
evident from comparing the two figures. The limit of this
representation takes place when the stable tube manifold allows for
collisions with the planet.  In Fig.~\ref{fig:tubes25} we present a
stable manifold that, while still folding clockwise, passes very close
to $P_2$. It is natural to expect that, by increasing the energy, the
stable manifold is such that some orbits eventually collide with
$P_2$. This is clearly visualized in Fig.~\ref{fig:tubes23} and,
particularly, in~\ref{fig:zooms}.  For the energy $E = -1.5113$, we
show that the manifold embraces the position of the planet, therefore
allowing possible collisions with it. In the right panel of
Fig.~\ref{fig:zooms} we show some of the orbits belonging to the
manifold, indicating with arrows the sense of the orbit when
integrated backward in time. We see that, besides passing extremely
close to the singularity, the orbits can either revolve around the
$P_2$ either clockwise or counter-clockwise, indicating that this
value of the energy is, already, a good representation of orbits of
type (iii) (see description in Introduction).


\section{Conclusions}

In this work, we studied the transition from non collisional to
possibly collisional orbits as a outcome of the change in the
structure of the tubes manifolds emanating from the Lyapunov orbits of
the CR3BP for the Sun-Jupiter system. To this aim, we constructed a
normal form of the Levi-Civita Hamiltonian, at a partially hyperbolic
fictitious equilibrium.  The normalization of the Levi-Civita
Hamiltonian, which is regular at the position of the planet, allows us
to reach values of the energy not investigated before via the
traditional normalizations of the CR3BP Hamiltonian in Cartesian
variables.  We found that, for suitably large values of the energy,
the tubes manifolds emanating from the horizontal Lyapunov orbits may
contain orbits that collide with the secondary body before performing
a full circulation around it (when integrated backwards from the
Lyapunov orbit), or can revolve either clockwise or counter-clockwise
around $P_2$. We established that, for $\mu = \mu_J$, the transition
takes place for values of the energy larger than $E > -1.5167$. We
expect this threshold to be relevant for future studies of dynamics of
temporary trapped comets and/or for space mission design.

As future work, we plan to extend the present work to
  the spatial restricted three--body problem. The Levi-Civita
  regularization is extended to the spatial case by the
  Kustaanheimo-Stiefel regularization,
  \begin{equation}
    \begin{aligned}
X &= u_1^2-u_2^2-u_3^2+u_4^2 \\
Y &= 2(u_2 u_2-u_3u_4)\\
Z &= 2( u_1 u_3+u_2u_4)\\
dt&= \norm{u}^2 d\tau
\end{aligned}
\end{equation}
see \cite{K64}, \cite{KS65}. Let us denote by ${\cal K}_E(u,U)$ the
Hamiltonian representing the Kustaanheimo-Stiefel regularization (see
for example \cite{CG19a}, \cite{CG19b} for a detailed
derivation). For values $E>E_{L_1}$, the fictitious equilibrium found
in the regularized planar problem, $u_*=(0, \ustar)$, $U_*=(-2
\ustar^3,0)$, extends to an equilibrium of Hamilton equations of the
Kustaanheimo-Stiefel Hamiltonian, $u_*=(0, \ustar,0,0)$, $U_*=(-2
\ustar,0,0,0)$, satisfying the bi-linear equation $\ell
(u_*,U_*)=0$ where $\ell(u,U)= u_4 U_1- u_3 U_2+u_2 U_3-u_1 U_4$.  In
principle, this be used to define a normalizing transformation as done
in the present work. On the other hand, the internal symmetry of the
Kustaanheimo-Stiefel transformation requires specific adaptations,
which will be the subject of future work.


\phantomsection
\section*{Acknowledgments} 

\addcontentsline{toc}{section}{Acknowledgments} 

R.P. was supported by the ERC project 677793 StableChaoticPlanetM
(01/07/2018-30/06/2019) and the project MIUR-PRIN 20178CJA2B
(01/07/2019-30/06/2020).  M.G. also acknowledges the project MIUR-PRIN
20178CJA2B "New frontiers of Celestial Mechanics: theory and
applications".

\phantomsection
\bibliographystyle{unsrt}
\bibliography{biblio}


\noindent
\section*{Appendix}
{\bf A1) Computing the equilibria $(u_*,U_*)$.} The equilibrium points  
of ${\cal K}_E$, satisfying $u_1=0$ (which is the condition 
granting $y=2u_1u_2=0$, $x=x_2+u_1^2-u_2^2<x_2$) 
and $u_2\in (-1,0)$ ($u_2=-1$, $u_2=0$ locate the equilibrium on the 
primary body $P_1$ and $P_2$ respectively) are the solutions of
{\small
\begin{equation}\label{eq:ap1}
  \begin{aligned}
  \left. \frac{\partial {\cal K}_E}{\partial U_1} \right|_{u_1=0}
  &= \frac{1}{4} \, (U_1+2u_2^3) = 0 ~~, \\
  \left. \frac{\partial {\cal K}_E}{\partial U_2} \right|_{u_1=0}
  &= \frac{1}{4} \, U_2 = 0 ~~, \\
  \left. \frac{\partial {\cal K}_E}{\partial u_1} \right|_{u_1=0}
  &= -\frac{1}{2} \, U_2 \, u_2^2 = 0 ~~,\\
  \left. \frac{\partial {\cal K}_E}{\partial u_2} \right|_{u_1=0}
  &=  u_2 \bigg[-2 \big (E+\frac{1}{2}(1-\mu)^2\big ) -3 u_2^4 \Big. \\
    & +\frac{3}{2} u_2^2 \, (U_1+2 u_2^3) \\
    & -(1-\mu)\, u_2 \, \big( -2 u_2 -
    \frac{-2 u_2+2u_2^3}{(1-2 u_2^2+u_2^4)^{3\over 2}} \big) \\
    & \Big. -2 (1-\mu)\big (-u_2^2+ \frac{1}{\sqrt{1-2
              u_2^2+u_2^4}}\big) \bigg]=0~~.
  \end{aligned}
  \end{equation}}
We first notice that the first equation of this system provides 
$U_1=-2u_2^3$, while the second and the third ones are solved if $U_2=0$. 
Then, by  using $U_1=-2u_2^3$ and noticing that  
since $u_2^2 \in (0,1)$ we have $\sqrt{1-2 u_2^2+u_2^4}=1-u_2^2$,  
the last equation is solved by $u_2$ satisfying $F(u_2,E)=0$, where 
the function $F$ is defined in (\ref{eq:fu2e}). 
\vskip 0.2 cm
\noindent
{\bf A2) Solutions of the equation $F(u_2,E)=0$.}
To solve equation
$F(u_2,E)=0$ we first re-write it in the form
\begin{displaymath}
  \begin{aligned}
    E  = & f(u_2,\mu) \\
    := & \text{\small $- {3\over 2} u_2^4+(1-\mu)\left (2 u_2^2 -{ u_2^2\over (1-u_2^2)^2}
-{1\over 1-u_2^2}\right ) -{(1-\mu)^2\over 2}$} ,
  \end{aligned}
  \end{displaymath}
>From standard calculus, we have
\begin{displaymath}
{\partial f\over \partial u_2}=-6 u_2^3 -4(1-\mu)u_2^3{3-3u_2^2+u_2^4
\over (1-u_2^2)^3}>0
\end{displaymath}
for all $\mu\in [0,1/2]$ and $u_2\in (-1,0)$. As a consequence there is 
a strictly monotone increasing function $E:=E(u_2)$ defined for 
$u_2\in (-1,0)$ such that $F(u_2,E(u_2))=0$. Since we know that the Lyapunov 
orbits exist only for $E\geq E_{1}$, we restrict the domain of $E(u_2)$ to
the interval $[-\sqrt{x-x_{L_1}},0)$ and we define the inverse $a(u_2)$ 
of $E(u_2)$:
\begin{displaymath}
\Big [E_{1}, E_*\Big )\longrightarrow  
[-\sqrt{x-x_{L_1}},0)
\end{displaymath}
\begin{displaymath}
u_2  \longmapsto (\ustar)  
\end{displaymath}
where $E_*$ has been defined in \eqref{Ecrit}. 
\vskip 0.2 cm
\noindent
{\bf A3) Linearization of the Hamilton equations of ${\cal K}_E$ at
  $u_*,U_*$.}  The Jacobian matrix of the Hamilton vector field of
${\cal K}_E$ at $(u_*,U_*)$ is
\begin{equation}\label{eq:matrixM}
  M = 
  \begin{pmatrix}
    0 & \frac{3 (\ustar)^2}{2} & \frac{1}{4} & 0 \\
    -\frac{(\ustar)^2}{2} & 0 & 0 & \frac{1}{4} \\
    \Gamma & 0 & 0 & \frac{(\ustar)^2}{2} \\
    0 & - 2 \Gamma + (\ustar)^4 & -\frac{3 (\ustar)^2}{2} & 0 
  \end{pmatrix}~,
\end{equation}
with
\begin{displaymath}
\Gamma = -(\ustar)^4 + 4 (\ustar)^2(1 -\mu) \left(1 - \frac{1}{(1- (\ustar)^2)^3}\right) .
\end{displaymath}
As a consequence, the matrix $M$ has four eigenvalues:
\begin{equation}\label{eq:eigensolved}
  \begin{aligned}
  \gamma & = \pm {1\over \sqrt{8}}\sqrt{ - \Gamma - 5 (\ustar)^4 \pm \sqrt{9 \Gamma^2- 22 (\ustar)^4\Gamma -15 (\ustar)^8}}\\
  & := \pm \sqrt{{A\pm \sqrt{B}}\over 8}~.
  \end{aligned}
\end{equation}
whose nature is determined by the values of the functions $A,B$. From
standard calculus, for all $\mu\in (0,1/2)$ and $u_2\in (\ustar,0)$
one proves: $\Gamma <0$, $B>0$, $B>\Gamma+5 (\ustar)^4$.  As a consequence,
the matrix $M$ has a couple of real eigenvalues $\pm \lambda$ and a
couple of purely imaginary numbers $\pm i \omega$. Moreover, for $a$
going to zero, both $\lambda,\omega$ tend to zero as (a constant
multiplying) $(\ustar)^2$.
   \vskip 0.2 cm
\noindent
{\bf A4) Diagonalization of the quadratic part of ${\cal K}_E$.} Let $
{v}_{i}$, $i=1,\ldots ,4$, the four eigenvectors of $M$; matrix $A$ is
constructed as :
\begin{equation}\label{eq:matrixA}
  A=
   ( c_1  {v}_{3}^t ,\, c_2  {v}_{1}^t ,\, \mathrm{i} c_1  {v}_{4}^t ,\, c_2  
{v}_{2}^t) 
\end{equation}
where $c_1,c_2 \in \mathbb{C}$ are chosen to satisfy 
the symplectic condition $A^{t}  {\mathbb J}  A = {\mathbb J}$,
where ${\mathbb J}$ denotes the standard symplectic matrix.


\end{document}